\documentclass[runningheads]{llncs}

\title{Generating faster algorithms for\\ $d$-\textsc{Path Vertex Cover}\\(full version)\thanks{The authors acknowledge the support of the OP VVV MEYS funded project CZ.02.1.01/0.0/0.0/16\_019/0000765 ``Research Center for Informatics'' and the Grant Agency of the CTU in Prague funded grant No. SGS20/208/OHK3/3T/18.}}

\titlerunning{Generating faster algorithms for\\ $d$-\textsc{Path Vertex Cover}}
\author{Radovan Červený\orcidID{0000-0003-4528-9525}  and Ondřej Suchý\orcidID{0000-0002-7236-8336} }

\authorrunning{R. Červený and O. Suchý}

\institute{Department of Theoretical Computer Science, Faculty of Information Technology,\\
Czech Technical University in Prague, Prague, Czech~Republic\\
\email{\{radovan.cerveny,ondrej.suchy\}@fit.cvut.cz}}

\usepackage[utf8]{inputenc}

\usepackage{float}
\usepackage{amsmath}
\usepackage{amssymb}
\usepackage{aliascnt}
\usepackage{caption}
\usepackage[list=true,listformat=simple]{subcaption}
\usepackage{algorithm}
\usepackage[noend]{algpseudocode}
\usepackage{hhline}
\usepackage{makecell}
\usepackage{multirow}
\usepackage{mathtools}
\usepackage[figuresright]{rotating}
\usepackage{diagbox}
\usepackage{pgfplots}
\usepackage{comment}
\usepackage{tabularx}
\usepackage[hyphens]{url}
\usepackage{hyperref}
\usepackage[hyphenbreaks]{breakurl}
\usepackage{lineno}
\usepackage{cleveref}
\usepackage{pgfplots}
\usepackage{etoolbox}
\usepackage{enumerate}
\pgfplotsset{compat=1.9}

\ifcsdef{observation}{}{
\newtheorem{observation}{Observation}
}
\newtheorem{rrule}{Reduction Rule}
\crefname{rrule}{Reduction Rule}{Reduction Rules}
\Crefname{rrule}{Reduction Rule}{Reduction Rules}

\newcommand{\FSVD}{\textsc{$\F$-SVD}}

\newcommand{\ostar}{\mathcal{O}^*}

\newcommand{\bumpy}{bumpy}
\newcommand{\bumpset}{bump-inducing}
\newcommand{\bff}{\mathit{bf}}

\newcommand{\A}{\mathcal{A}}
\newcommand{\B}{\mathcal{B}}
\newcommand{\F}{\mathcal{F}}
\newcommand{\N}{\mathbb{N}}
\newcommand{\Z}{\mathbb{Z}}

\newcommand{\Llist}{\mathcal{L}}

\newcommand{\gLlist}{L}
\newcommand{\gLgood}{\gLlist_{\textit{good}}}
\newcommand{\gLbad}{\gLlist_{\textit{bad}}}

\DeclarePairedDelimiter\dofloor{\lfloor}{\rfloor}

\usetikzlibrary{backgrounds}
\usetikzlibrary{calc}
\usetikzlibrary{decorations.pathreplacing}

\definecolor{HighlightColor}{RGB}{255,255,0}
\definecolor{HighlightColor2}{RGB}{100,255,100}
\definecolor{HighlightColorSupergraph}{RGB}{100,255,100}
\definecolor{DimColorLighter}{RGB}{245,245,245}
\definecolor{DimColor}{RGB}{200,200,200}
\definecolor{DimColorDarker}{RGB}{120,120,120}

\tikzset{HighlightPath/.style={draw, line width=10pt, HighlightColor, cap=round, rounded corners=1pt}}
\tikzset{DimNode/.style={fill=DimColorLighter,circle,inner sep=1pt,draw=DimColor}}
\tikzset{DimPath/.style={draw, dashed, DimColor}}

\tikzset{NormalNode/.style={fill=black,circle,inner sep=1pt}}
\tikzset{RedNode/.style={fill=red,circle,inner sep=2pt}}

\usepackage{etoolbox}

\newcommand{\lv}[1]{}

\newcommand{\appendixText}{}

\begin{document}
\maketitle

\begin{abstract}
Many algorithms which exactly solve hard problems require branching on more or less complex structures in order to do their job. Those who design such algorithms often find themselves doing a meticulous analysis of numerous different cases in order to identify these structures and design suitable branching rules, all done by hand. This process tends to be error prone and often the resulting algorithm may be difficult to implement in practice. 

In this work, we aim to automate a part of this process and focus on the simplicity of the resulting implementation. 

We showcase our approach on the following problem.
For a constant $d$, the $d$-\textsc{Path Vertex Cover} problem ($d$-PVC) is as follows: Given an undirected graph and an integer $k$, find a subset of at most $k$ vertices of the graph, such that their deletion results in a graph not containing a path on $d$ vertices as a subgraph.
We develop a fully automated framework to generate parameterized branching algorithms for the problem and obtain algorithms outperforming those previously known for $3 \le d \le 8$, e.g., we show that $5$-PVC can be solved in $O(2.7^k\cdot n^{O(1)})$ time.

\end{abstract}

\section{Introduction}
The motivation behind this paper is to renew the interest in computer aided design of graph algorithms which was initiated by Gramm et al.~\cite{GrammGHN04}. Many parameterized branching algorithms follow roughly the same pattern: 1) perform a meticulous case analysis; 2) based on the analysis, construct branching and reduction rules; 3) argue that once the rules cannot be applied, some specific structure is achieved. Also, depending on how ``deeply'' you perform the case analysis, you may slightly improve the running time of the algorithm, but bring nothing new to the table.

This paper aims to provide a framework which could help in the first two steps of the pattern at least for some problems.
We phrase the framework for a rather general problem which is as follows. For any nonempty finite set of connected graphs $\mathcal{F}$ we define the problem \textsc{$\mathcal{F}$-Subgraph Vertex Deletion, $\mathcal{F}$-SVD,} where, given a graph $G=(V,E)$ and an integer~$k$, the task is to decide whether there is a~subset~$S$ of at most $k$ vertices of $G$ such that $G \setminus S$ does not contain any graph from $\mathcal{F}$ as a subgraph (not even as a non-induced one).
While we only apply the framework to the problem of \textsc{$d$-PVC} defined later, the advantage of phrasing the framework for $\mathcal{F}$-SVD is twofold. First, it makes it easier to apply it to other problems. Second, the general notation introduced makes the description less cluttered.

Since the problem is NP-complete for most reasonable choices of $\mathcal{F}$, as follows from the meta-theorem of Lewis and Yannakakis~\cite{LewisY80}, any algorithm solving the problem exactly is expected to have exponential running time.
In this paper we aim on the parameterized analysis of the problem, that is, to confine the exponential part of the running time to a specific parameter of the input, presumably much smaller than the input size.
In particular, we only use the most standard parameter, which is the desired size of the solution $k$, also called \emph{the budget}.
Algorithms achieving running time $f(k)n^{O(1)}$ are called \emph{parameterized}, \emph{fixed-parameter tractable}, or \emph{FPT}. See Cygan et al.~\cite{CyganFKLMPPS15} for a broader introduction to parameterized algorithms.

To understand how parameterized branching algorithms typically work, consider the following simple recursive algorithm for $\mathcal{F}$-SVD.
We find in the input graph $G$ an occurrence $F'$ of graph $F$ from $\mathcal{F}$.
We know that at least one of the vertices of $F'$ must be in any solution.
Hence, for each vertex of $F'$ we try adding it to a prospective solution, decreasing the remaining budget by one, and recursing. The recursion is stopped when the budget is exhausted, or there are no more occurrences of graphs from $\mathcal{F}$ in $G$, i.e., we found a solution.
It is easy to analyze that this algorithm has running time\footnote{The $\ostar()$ notation suppresses all factors polynomial in the input size.} $\ostar(d^k)$, where $d$ is the number of vertices of the largest graph in $\mathcal{F}$.
Many parameterized branching problems follow a similar scheme, branching into a constant number of alternatives in each step, for each alternative making a recursive call with the budget (or some other parameter) decreased by some constant.

One can improve upon this trivial algorithm by looking at $F'$ together with its surroundings. Working with this larger graph $F''$ often allows for more efficient branching as now multiple \emph{overlapping} occurences of graphs from $\mathcal{F}$ may appear in $F''$ instead of just one.
Our framework and that of Gramm et al.~\cite{GrammGHN04} rely upon this observation, as they iteratively take larger and larger graphs into consideration---similarly to what a human would do, but on a much larger scale.

The fundamental novelty of our framework in comparison to that of Gramm et al.~\cite{GrammGHN04} is that we are able to identify which vertices of the graph $F''$ under consideration can still have outside neighbors and which do not. We call the latter ``red''. This way we are able to say that if you find an occurrence of $F''$ in the input graph, you can be sure that the red vertices do not have neighbors in the input graph apart from those that are in $F''$.

This additional information allows us to eliminate some branches of the constructed branching rules, rapidly improving their efficiency. It also reduces the number of graphs we need to consider and also allows us to design better reduction rules to aid our framework.

We apply the general framework to the problem of \textsc{$d$-Path Vertex Cover ($d$-PVC)}.
The problem lies in determining a~subset~$S$ of vertices of a~given graph $G=(V,E)$ of at most a given size $k$ such that $G \setminus S$ does not contain a~path on~$d$ vertices (even not a non-induced one). It was first introduced by Brešar et al.~\cite{BresarKKS11}, but its NP-completeness for any $d \ge 2$ follows already from the above-mentioned meta-theorem of Lewis and Yannakakis~\cite{LewisY80}. The \textsc{2-PVC} problem corresponds to the well known \textsc{Vertex Cover} problem and the \textsc{3-PVC} problem is also known as \textsc{Maximum Dissociation Set} or \textsc{Bounded Degree-One Deletion}. The \textsc{$d$-PVC} problem is motivated by the field of designing secure wireless communication protocols~\cite{Novotny10} or route planning and speeding up shortest path queries~\cite{FunkeNS16}.

As mentioned above, \textsc{$d$-PVC} is directly solvable by a~trivial FPT algorithm that runs in $\ostar(d^k)$ time. However, since \textsc{$d$-PVC} is a special case of \textsc{$d$-Hitting Set}, it follows from the results of Fomin et al.~\cite{FominGKLS10} that for any $d\ge4$ we have an algorithm solving \textsc{$d$-PVC} in $\ostar((d - 0.9245)^{k})$ time. For $d \ge 6$ algorithms with even better running times are presented in the work of Fernau~\cite{Fernau10}.

In order to find more efficient solutions, the problem has been extensively studied in a~setting where~$d$ is a~small constant. This is in particular the case for the \textsc{2-PVC} (\textsc{Vertex Cover}) problem~\cite{BalasubramanianFR98,BussG93,ChandranG05,ChenKJ01,ChenLJ00,DowneyF92,NiedermeierR99,NiedermeierR03}, where the algorithm of Chen, Kanj, and Xia~\cite{ChenKX10} for a long time held the best known running time of $\ostar(1.2738^k)$,
but recently Harris and Narayanaswamy~\cite{HarrNaraFastVC} claimed the running time of $\ostar(1.25288^k)$. For \textsc{3-PVC}, Tu~\cite{Tu15} used iterative compression to achieve a~running time of $\ostar(2^k)$.
This was later improved by Katrenič~\cite{Katrenic16} to $\ostar(1.8127^k)$, by Xiao and Kou~\cite{XiaoK17} to $\ostar(1.7485^k)$ by using a~branch-and-reduce approach and finally by Tsur~\cite{Tsur19a} to $\ostar(1.713^k)$. For the \textsc{4-PVC} problem, Tu and Jin~\cite{TuJ16} again used iterative compression and achieved a~running time of $\ostar(3^k)$ and Tsur~\cite{Tsur21-4PVC} gave the current best algorithm that runs in $\ostar(2.619^k)$ time. The authors of this paper developed an $\ostar(4^k)$ algorithm for \textsc{$5$-PVC}~\cite{CervenyS19}. For $d=5$, $6$, and $7$ Tsur~\cite{Tsur23} discovered algorithms for $d$-PVC with running times $\ostar(3.945^k)$, $\ostar(4.947^k)$, and $\ostar(5.951^k)$, respectively.


Using our automated framework, we are able to present algorithms with improved running times for some \textsc{$d$-PVC} problems when parameterized by the size of the solution $k$. The results are summarized in \autoref{results-summary}.

\begin{table}[t]
\centering
\caption[Improved running times of some \textsc{$d$-PVC} problems.]{Improved running times of some \textsc{$d$-PVC} problems.}
\label{results-summary}
\begin{tabular}{l|l|l|r}
\textsc{$d$-PVC} & Previously known & Our result & Our \# of rules\\\hline
\textsc{$2$-PVC} & $\ostar(1.25288^k)$ \cite{HarrNaraFastVC} & $\ostar(1.3294^k)$ & 9,345,243 \\\hline
\textsc{$3$-PVC} & $\ostar(1.713^k)$  \cite{Tsur19a}     & $\ostar(1.708^k)$  & 1,226,384 \\\hline
\textsc{$4$-PVC} & $\ostar(2.619^k)$  \cite{Tsur21-4PVC} & $\ostar(2.138^k)$  & 911,193 \\\hline
\textsc{$5$-PVC} & $\ostar(3.945^k)$  \cite{Tsur23}     & $\ostar(2.636^k)$  & 739,542 \\\hline
\textsc{$6$-PVC} & $\ostar(4.947^k)$  \cite{Tsur23}     & $\ostar(3.334^k)$  & 414,247 \\\hline
\textsc{$7$-PVC} & $\ostar(5.951^k)$  \cite{Tsur23}     & $\ostar(3.959^k)$  & 5,916,297 \\\hline
\textsc{$8$-PVC} & $\ostar(7.0237^k)$ \cite{Fernau10}    & $\ostar(5.654^k)$  & 296,044 \\\hline
\end{tabular}
\end{table}

\paragraph*{Further Related Work}
The only other approach to generating algorithms with provable worst-case running time upper bounds we are aware of is limited to algorithms for SAT~\cite{FedinK04,KojevnikovK06,Kulikov05}.

Similar methods as for parameterized branching algorithms are often used for moderately exponential algorithms.
Here one measures the running time solely in terms of the input size.
Several efficient (faster than trivial enumeration) exact algorithms are known for \textsc{2-PVC} and \textsc{3-PVC}. In particular, \textsc{2-PVC} (\textsc{Vertex Cover}) can be solved in $\mathcal{O}(1.1996^n)$ time and polynomial space due to Xiao and Nagamochi~\cite{XiaoN17} and \textsc{3-PVC} can be solved in $\mathcal{O}(1.4613^n)$ time and polynomial space due to Chang et al.~\cite{ChangCHLRS18} or in $\mathcal{O}(1.3659^n)$ time and exponential space due to Xiao and Kou~\cite{XiaoK17exact}.

\paragraph*{Organization of the Paper}
In \autoref{sec:prelim} we specify what branching rules we head for, how to apply them, when they are correct and similar fundamental notions.
In \autoref{sec:output} we explain how a good set of branching rules can be turned into a correct algorithm for $\mathcal{F}$-SVD.
\autoref{sec:FAB_algo} describes the algorithm for generating a good set of rules, whereas \autoref{generate-function-section} describes the way in which we create a single correct branching rule with good branching factors.
The specifics of applying the framework to $d$-PVC and the results obtained for this problem are described in \autoref{sec:applying}.
\autoref{sec:proofs} describes the data we provide for the generated algorithms and the way to use them to prepare an efficient implementation of the algorithms.
We conclude the paper with some ideas for future research in \autoref{sec:conclusions}.


\section{Fundamental Definitions and Basic Observations}\label{sec:prelim}
In this paper we are going to assume that vertex sets of all graphs are finite subsets of $\N$, the set of all non-negative integers, i.e., we have a set of all graphs. Furthermore, when adding a graph into a set of graphs, we only add the graph if none of the graphs already in the set is isomorphic to it. Similarly, when forming a set of graphs we only add one representative for each isomorphism class. Finally, when subtracting a graph from a set, we remove from the set all graphs isomorphic to it.

For any nonempty finite set of connected graphs $\F$ we define the problem:

\vspace{2mm}
\noindent
\begin{tabularx}{\textwidth}{|l|X|}
  \hline
\multicolumn{2}{|l|}{\textsc{$\mathcal{F}$-Subgraph Vertex Deletion, $\mathcal{F}$-SVD}} \\ \hline
  \textsc{Input}: & A~graph $G=(V,E)$, an integer $k \in \N$. \\
  \textsc{Output}: & A~set $S \subseteq V$, such that $|S| \leq k$ and no subgraph of $G \setminus S$ is isomorphic to a graph in $\F$. \\
  \hline
\end{tabularx}
\vspace{2mm}

We call $\F$ of {\FSVD} a \emph{\bumpset} set.
We call a graph $G$ \emph{\bumpy} if it contains some graph from the {\bumpset} set $\F$ as a subgraph.
We call a vertex subset $S$ a \emph{solution} (for a graph $G=(V,E)$), if the graph $G \setminus S$ is not \bumpy. Since $\F$ is finite, checking if $G$ is {\bumpy} is polynomial in the size of $G$.

Next we define a variant of a supergraph with a restriction that the original graph has to be an induced subgraph of the supergraph.

\begin{definition}[expansion, $i$-expansion, $\sigma$, $\sigma_i$, $\sigma^*$]
Let $H$ be a connected graph.
A graph $G$ is an \emph{expansion} of $H$, if $G$ is connected, $V(H) \subseteq V(G)$ and $G[V(H)]=H$.
It is an $i$-\emph{expansion} for $i \in \N$ if furthermore $|V(G)|=|V(H)|+i$.
For $i \in \N$ let $\sigma_i(H)$ denote the set of all $i$-expansions of $H$ (note again that we take only one representative for each isomorphism class). As shorthand, we will use $\sigma(H) = \sigma_1(H)$. Let $\sigma^*(H) = \bigcup_{i \in \mathbb{N}} \sigma_i(H)$ denote the set of all expansions of~$H$.
\end{definition}

A central notion to our approach is the following (restricted) variant of a branching rule.

\begin{definition}[Subgraph branching rule]
A \emph{subgraph branching rule} is a triple $(H,R,\B)$, where $H$ is a connected {\bumpy} graph, $R \subseteq V(H)$ is a set of \emph{red} vertices (representing the vertices supposed not to have neighbors outside $H$), and $\B \subseteq \big(2^{V(H)}\setminus \{\emptyset\}\big)$ is a non-empty set of \emph{branches}.
\end{definition}

\begin{definition}[An application of a subgraph branching rule]
We say that a subgraph branching rule $(H,R,\B)$ \emph{applies} to graph $G$, if $G$ contains an induced subgraph $H'$ isomorphic to $H$ by isomorphism $\phi: V(H) \to V(H')$ \emph{(witnessing isomorphism)} and for every $r \in R$ we have $N_G(\phi(r)) \subseteq V(H')$.
In other words, the vertices of $H'$ corresponding to red vertices only have neighbors inside the subgraph $H'$.
If the rule applies and the current instance is $(G,k)$, then the algorithm makes for each $B \in \B$ a recursive call with instance $(G \setminus \phi(B),k-|B|)$.
\end{definition}

Note that we do not allow $\emptyset \in \B$. Therefore the budget gets reduced and we are making progress in every branch.

\begin{definition}[Correctness of a subgraph branching rule]
A subgraph branching rule $(H,R,\B)$ is \emph{correct}, if for every $G$ and every solution $S$ for~$G$ such that $(H,R,\B)$ applies to $G$ and $\phi: V(H) \to V(H')$ is the witnessing isomorphism, there exists a solution $S'$ for $G$ with $|S'| \le |S|$ and a branch $B \in \B$ such that $\phi(B) \subseteq S'$.
\end{definition}

\begin{definition}[Branching factor of a subgraph branching rule]
For any subgraph branching rule $(H, R, \B)$ let $\bff((H, R, \B))$ be the branching factor of the branches in $\B$, i.e., the unique positive real solution of the equation: $1 = \sum_{B \in \B} x^{-|B|}$ (see \cite[Chapter 2.1 and Theorem 2.1]{FominK10} for more information on (computing) branching factors).
\end{definition}

If $|\B|=1$, i.e., there is exactly one branch, then such a rule is rather a~reduction rule than a~branching rule. In particular, the above equation is only satisfied with $x=1$ and the branching factor of such a rule is thus $1$ according to our definition. Indeed, if we only applied such rules, then the running time of the algorithm would be polynomial, i.e., $O^*(1^k)$. To simplify the description, we will treat these rules as all other subgraph branching rules.

\begin{observation}\label{correct-subgraph-branching-rule-existence-observation}
For any connected \bumpy{} graph $H$ and any $R \subseteq V(H)$ we can always construct at least one correct subgraph branching rule.
\end{observation}

 \begin{proof}
Let $G$ be a connected \bumpy{} graph. Consider the rule $\varrho = (H, R, \B)$ where $\B = \{\{v\}\mid v \in H\}$ and assume that it applies to $G$ and $\phi$ is the witnessing isomorphism. Let $H'$ be the induced subgraph of $G$ to which $\varrho$ applies. As the graph $H$ is \bumpy{}, graph $H'$ must be also \bumpy{}. Thus for any solution $S$ for $G$ we have $S \cap V(H') \neq \emptyset$. Since each branch of the rule $\varrho$ contains one single vertex of $H$, for at least one of the branches $\{v\} \in \B$ we have $\phi(v) \in S \cap V(H')$. Therefore the rule $\varrho$ is correct.
\qed
\end{proof}

The following definition formalizes a function that, given a graph $H$ and a set of vertices~$R$, computes a set $\B$ of branches such that $(H, R, \B)$ is a correct subgraph branching rule.

\begin{definition}[Brancher]
A \emph{brancher} is a function which assigns to any connected \bumpy{} graph $H$ and $R \subseteq  V(H)$
a correct branching rule $\tau(H, R)= (H, R, \B)$ for some non-empty set $\B \subseteq \big(2^{V(H)}\setminus \{\emptyset\}\big)$.
For a brancher $\tau$ as a shorthand let $\tau(H) = \tau(H, \emptyset)$. For a set of graphs $\{H_1, H_2, \ldots, H_r\}$ we will have $\tau(\{H_1, H_2, \ldots, H_r\}) = \{\tau(H_1), \tau(H_2), \ldots, \tau(H_r)\}$.
\end{definition}

The above observation shows that at least one brancher exists.


Next we summarize some basic operations with lists that we are going to use.

\begin{definition}
Let $L=(e_1, e_2, \ldots, e_r)$ be an ordered list of elements of any type. The type will always be specified or derivable from the context.

For two lists $L_1 = (e_1, e_2, \ldots, e_r)$ and $L_2 = (e'_1, e'_2, \ldots, e'_{r'})$ a \emph{concatenation} of lists $L_1$ and $L_2$, denoted $L_1 \circ L_2$ is the list $L_1 \circ L_2 = (e''_1, e''_2, \ldots, e''_{r+r'})$ where $e''_i = e_i, i\in\{1, 2, \ldots, r\}$ and $e''_{r+j} = e'_j, j \in \{1, 2, \ldots, r'\}$.

To remove an element $e_i$ from list $L = (e_1, e_2, \ldots, e_i, \ldots, e_r\}$ we will use the notation $L \setminus e_i$, i.e., $L \setminus e_i = (e_1, e_2, \ldots, e_{i-1}, e_{i+1}, \ldots, e_r)$.

Whenever we use a set in a place where an ordered list is expected, we construct an ordered list from the set choosing an arbitrary order. Whenever we use an ordered list in a place where a set is expected, we naturally throw away the ordering.

\end{definition}

\begin{definition}
For a set of subgraph branching rules $\Llist = (\varrho_1, \varrho_2, \ldots, \varrho_r)$ where $\varrho_i = (H_i, R_i, \B_i)$ we will denote $\Psi(\Llist) = \max \{|V(H_i)| \mid (H_i, R_i, \B_i) \in \Llist\}$ the maximum number of vertices among the graphs of the subgraph branching rules in $\Llist$.
\end{definition}

In our algorithms, apart from rules generated by the generating algorithm, we will be using some handmade reduction\footnote{Roughly speaking, a reduction rule is a polynomial-time procedure that replaces the input instance with another one, preserving the answer.} or branching rules. The rules help the algorithm and allow to steer it away from some difficult corner cases. Typically, these rules ensure that some substructures do not appear in the input graph any more. We will denote the set of the handmade rules $\A$.

Next we define the crucial property of a set of subgraph branching rules which forms a base for the proof of correctness of the generated algorithm.

\begin{definition}
A set of subgraph branching rules $\Llist = (\varrho_1, \varrho_2, \ldots, \varrho_r)$ is called \emph{exhaustive with respect to $\A$} if every rule $\varrho_i$ is correct and for every connected \bumpy{} graph $G$ to which no handmade rule in $\A$ applies and which has at least $\Psi(\Llist)$ vertices there is a subgraph branching rule $\varrho_i$ in $\Llist$ that applies to $G$.
If the set is exhaustive with respect to $\emptyset$, that is, even without any handmade rules, we will omit the ``with respect to $\A$'' clause.
\end{definition}

During all our operations we aim to maintain an exhaustive set of subgraph branching rules.

The following observation identifies our starting set of graphs. 

\begin{observation}\label{initial-list-observation}
Let $\mathcal{F}$ be the {\bumpset} set of some {\FSVD} problem. Let $f = \max_{H \in \mathcal{F}}{|V(H)|}$. Let $L = \{F_1, F_2, \ldots, F_r\}$ be the set of all connected \bumpy{} graphs with $f$ vertices. Let $\tau$ be a brancher. Then the set of subgraph branching rules $\Llist = \tau(L)$ is exhaustive.
\end{observation}
 \begin{proof}
Let $G$ be a connected \bumpy{} graph which has at least $\Psi(\Llist)=f$ vertices.
As $G$ is \bumpy{}, connected and has at least $f$ vertices, there exists an induced subgraph $G'$ of $G$ such that $G'$ is \bumpy{}, connected and has \emph{exactly} $f$ vertices.
Since the set $L$ contains all connected \bumpy{} graphs with $f$ vertices, one of them, $F_i$, is isomorphic to $G'$.
As $\Llist$ was constructed from the set $L$, the rule $\tau(F_i) = \tau(F_i, \emptyset)$ from $\Llist$  applies to $G$.
Because all the rules in $\Llist$ are correct by definition, we have that the set $\Llist$ is exhaustive.
\qed
\end{proof}

\section{The Output Algorithm and Its Correctness}
\label{sec:output}

Our goal will be to obtain a set $\Llist$ of subgraph branching rules with good branching factors which is exhaustive with respect to $\A$.
This section summarizes how we use the set to design an algorithm for \textsc{$\F$-SVD} once we obtain such a set.
We call the algorithm $(\A,\Llist)$-Algorithm for \textsc{$\F$-SVD} and its pseudocode is available in Algorithm \ref{fab-output-algorithm-pseudocode}.

\begin{algorithm}[t!]
\caption{$(\A,\Llist)$-Algorithm for \textsc{$\F$-SVD}}\label{fab-output-algorithm-pseudocode}
\begin{algorithmic}[1]
 \State Let $\A$ be a list of handmade rules and $\Llist$ be a set of subgraph branching rules.
 \Function{$\mathit{SolveRecursively}$}{$G,k$}
 \If{$k < 0$} \label{ok1-fab-output-algorithm-pseudocode}
   \State Return NO.
 \EndIf
 \If{$G$ is not \bumpy{}} \label{output_algo:not_bumpy}
    \State Return YES.
 \EndIf
 \If{$k = 0$}
   \State Return NO.
 \EndIf \label{step:last_stopping}
 \If{Some rule from $\A$ can be applied to $G$} \label{okA-fab-output-algorithm-pseudocode}
   \State Find the first rule $\varrho_A$ from $\A$ that can be applied to $G$. Apply $\varrho_A$ to $G$ and return the corresponding answer (might involve recursive calls to \textit{SolveRecursively}).
 \EndIf \label{okAend-fab-output-algorithm-pseudocode}
 \If{Each \bumpy{} connected component of $G$ has less than $\Psi(\Llist)$ vertices} \label{okBrute-fab-output-algorithm-pseudocode}
    \State Find the optimal solution for each component separately by brute-force.
    \State Let the solutions be $S_1, S_2, \ldots, S_c$.
    \If{$\sum_{i=1}^{c}|S_i| \leq k$}
      \State Return YES.
    \Else
      \State Return NO.
    \EndIf
 \EndIf \label{ok2-fab-output-algorithm-pseudocode} \label{okBruteEnd-fab-output-algorithm-pseudocode}
 \State Let $C$ be the vertices of the \bumpy{} connected component of $G$ with at least $\Psi(\Llist)$ vertices. \label{ok3-fab-output-algorithm-pseudocode}
 \State Find a branching rule $(H,R,\B)$ from the set $\Llist$ that can be applied to $G[C]$.
 \State Let $\phi$ be the corresponding isomorphism.
 \For {$B \in \B$}
   \If{\textit{SolveRecursively}($G \setminus \phi(B),k - |B|$) outputs YES}
     \State Return YES.
   \EndIf
 \EndFor
 \State Return NO.\label{faboutput_end}
 \EndFunction
\end{algorithmic}
\end{algorithm}

The algorithm first applies some trivial stopping conditions (lines \ref{ok1-fab-output-algorithm-pseudocode} to \ref{step:last_stopping}).
Then it applies the rules from $\A$ (lines \ref{okA-fab-output-algorithm-pseudocode} to \ref{okAend-fab-output-algorithm-pseudocode}).
Next, if every connected component is small, it finds a solution for each of them separately by a brute force (lines \ref{okBrute-fab-output-algorithm-pseudocode} to \ref{okBruteEnd-fab-output-algorithm-pseudocode}).
Finally, it takes a component which is large enough and finds a subgraph branching rule from $\Llist$ that applies to the component and applies it by making the appropriate recursive calls (lines \ref{ok3-fab-output-algorithm-pseudocode} to \ref{faboutput_end}).

The following theorem states that this algorithm is indeed correct.

\begin{theorem}
\label{thm:output_algo_correct}
Let $\A$ be a list of handmade rules and $\Llist$ be a set of subgraph branching rules.
If $\Llist$ is exhaustive with respect to $\A$, all rules in $\A$ are correct and can be applied in polynomial time, and each branching rule in $\A \cup \Llist$ has branching factor at most $\beta$, then
the $(\A,\Llist)$-Algorithm for \textsc{$\F$-SVD} is correct and runs in $\ostar(\beta^k)$ time.
\end{theorem}

\begin{proof}
We prove the correctness by induction on the height of the recursion tree.
If the answer is obtained directly by the recursion stopping conditions on lines \ref{ok1-fab-output-algorithm-pseudocode} to \ref{step:last_stopping} then it is correct.
The correctness of an answer from lines \ref{okA-fab-output-algorithm-pseudocode} to \ref{okAend-fab-output-algorithm-pseudocode} follows from the correctness of the handmade rules in $\A$ and, possibly, by the induction hypothesis.
Since all graphs in $\F$ are connected, lines \ref{okBrute-fab-output-algorithm-pseudocode} to \ref{okBruteEnd-fab-output-algorithm-pseudocode} are also correct.

If the computation reaches line \ref{ok3-fab-output-algorithm-pseudocode}, as no handmade rule from $\A$ can be applied to $G$ and the graph $G[C]$ has at least $\Psi(\Llist)$ vertices, there must be a rule in $\Llist$ that can be applied to $G[C]$ as the set $\Llist$ is exhaustive.
Let $\varrho=(H,R,\B)$ be a rule that applies and $\phi$ be the witnessing isomorphism.

On one hand, if for any $B \in \B$ the call \textit{SolveRecursively}($G \setminus \phi(B),k - |B|$) outputs YES, then, by the induction hypothesis, there is a solution $S'$ for $G \setminus \phi(B)$ of size at most $k - |B|$. Then, however, $S= S' \cup \phi(B)$ is of size at most $k$ and a solution for $G$, since $G \setminus S = (G\setminus \phi(B)) \setminus S'$. Therefore the answer YES is correct in this case.

On the other hand, if $S$ is a solution for $G$ of size at most $k$, then, since $\varrho$ is correct, there is a solution $S'$ for $G$ and $B \in \B$ such that $\phi(B) \subseteq S'$ and $|S'|\le |S|\le k$.
Then $S''= S' \setminus \phi(B)$ is of size at most $k - |B|$. Furthermore, it is a solution for $G \setminus \phi(B)$, as $(G \setminus \phi(B)) \setminus S''= G \setminus (\phi(B) \cup S'') = G \setminus S'$ and $S'$ is a solution for $G$. Therefore, in this case the call \textit{SolveRecursively}($G \setminus \phi(B),k - |B|$) will answer YES by the induction hypothesis and hence also this call will answer correctly.

It is assumed that the rules in $\A$ can be applied in polynomial time and have branching factors at most $\beta$. The optimal solution of each \bumpy{} connected component of $G$ with less than $\Psi(\Llist)$ vertices can be computed in polynomial time. Finally, all the rules in $\Llist$ can be applied in polynomial time (by testing all possible injections of $H$ in $V(C)$) and have branching factors at most $\beta$. Therefore, the running time is $\ostar(\beta^k)$.
\qed
\end{proof}

\paragraph*{Implementation Considerations}

Let us now discuss the effort needed to implement such an algorithm.
As we consider generating the set $\Llist$ by a computer program, we can safely assume that the set is available in a machine-readable format.
Then, to implement the algorithm, apart from implementing the rules from $\A$ and a brute-force solution for the small components, one has to implement the test on line~\ref{output_algo:not_bumpy} and a subprocedure taking care about application of the rules.
For the former, without any further knowledge of the particular {\FSVD} problem, one would probably use some algorithm for \textsc{Subgraph Isomorphism} to test whether some of the graphs in $\F$ is a subgraph of the input graph $G$ (see, e.g., \cite{MarxP14} for a survey).
For the latter task, one can take the subgraph branching rules from $\Llist$ one by one and for each of them test whether it applies to $C$ or not.
To decide whether $(H,R,\B)$ applies, one needs a \textsc{Subgraph Isomorphism} algorithm capable of extracting a witness and testing the condition on $R$.
Here the trivial algorithm which is enough for the proof could be prohibitively slow in practice.

To reduce the dependence of the running time of the output algorithm on the input size, one should consider using a kernelization algorithm as a preprocessing of the instance, which reduces the size of the input instance to polynomial in~$k$. The existence of such a kernelization for every \FSVD{} problem can be derived from (some) kernelizations for $d$-\textsc{Hitting Set}~\cite{FafianieK15}, the kernels for \textsc{$d$-PVC} are specifically considered in~\cite{CervenyCS2022}. For a more detailed account on kernelization algorithms see~\cite{FominLSZ2019}.

Nevertheless, we want to stress that the effort needed to implement the algorithm does not grow with the number of the rules in $\Llist$.
This might be considered an advantage over implementations of algorithms with many hand-made rules which are only described in a paper.

There are ways to reduce the (negative) influence of the number of rules on the running time of the algorithm.
We discuss some of them later after we describe the algorithm generating the rules.
Now we only mention two of them.

Often also the hand-made rules in $\A$ only produce an induced subgraph of the input graph.
In particular, all hand-made rules that we use are such. 
Then, a rule that does not apply to the current graph will never apply to the graph in any recursive calls made from the current call.
This allows us to cycle over the set of rules only once along each path of the recursion.

Furthermore, although we do not prove it formally, if the rules are used in the order as generated by our algorithm, then the neighborhood check for vertices in $R$ is always satisfied, i.e., unnecessary. One can then use faster \textsc{Subgraph Isomorphism} algorithms to test whether the rule applies.

\section{$(\F,\A,\beta)$-Algorithm}\label{sec:FAB_algo}
In this section we describe the algorithm to generate a suitable list of subgraph branching rules.

For a fixed {\FSVD} problem the input of the algorithm are the \bumpset{} set $\F$, a function $\mathit{Handled}_{\A}$ which can identify the situations handled by the handmade branching and reduction rules in $\A$, and the target branching factor $\beta \in \mathbb{R}$. We assume that the handmade rules in $\A$ are correct in the context of the given {\FSVD} problem, they can be applied in polynomial time, and that the branching rules have branching factors at most~$\beta$.

The output of the algorithm is an ordered list of subgraph branching rules $\Llist$, exhaustive with respect to $\A$, such that every rule in $\Llist$ has branching factor at most $\beta$.

Note that the output $\Llist$ of the $(\F,\A,\beta)$-Algorithm satisfies the assumptions of Theorem~\ref{thm:output_algo_correct}.

\subsection{Overview of the algorithm}

The algorithm maintains an ordered list and a set of connected \bumpy{} graphs named $\gLgood$ and $\gLbad$, respectively. 
The list $\gLgood$ stores graphs that already give rise to good subgraph branching rules, whereas the set $\gLbad$ represents the substructures for which the algorithm did not find any effective way to tackle them yet.
The algorithm starts with $\gLgood$ empty and $\gLbad$ being the set from Observation \ref{initial-list-observation}.
Then it gradually shifts graphs from $\gLbad$ to $\gLgood$ and replaces the graphs in $\gLbad$ maintaining the invariant that $\tau(\gLgood \cup \gLbad)$ is an exhaustive set of subgraph branching rules (for any brancher $\tau$).
The algorithm stops when it succeeds to make $\gLbad$ empty.

The algorithm in each round tries to move as many graphs currently in $\gLbad$ to $\gLgood$.
To this end, it first ``colors'' the vertices of the graph based on the substructures already handled by the rules obtained from the graphs already in $\gLgood$ (function \textit{Color} described below takes care of that).
Having a graph with as many vertices colored in red as possible, it then tries to design a subgraph branching rule with the smallest branching factor out of it.
The function \textit{Generate} described below (and mainly in the next section) takes care of that, i.e., it is our brancher.
It also reveals if the substructure is already handled by a rule in $\A$ (hand-made rule).
If the branching factor of the produced rule is at most $\beta$, then the graph is moved to the end of $\gLgood$.
The algorithm repeats the above steps as long as possible.
Once no graph from $\gLbad$ can be moved to $\gLgood$ this way, the algorithm replaces all graphs in $\gLbad$ by all their 1-expansions and starts a next round.
This corresponds to deepening the analysis, i.e., considering larger parts of the input graph at once.
The function \textit{Expand} takes care of that.


The descriptions of the functions follow together with their key properties.

\subsection{\textit{Color} function}\label{subs:color}

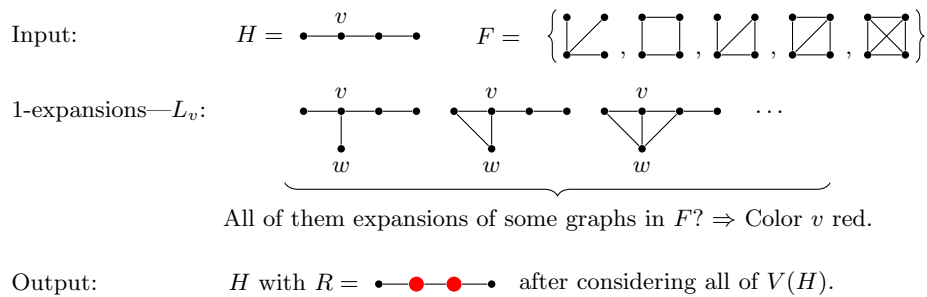
\begin{figure}
 \begin{tikzpicture}

\begin{scope}[shift={(0,0)}]
\path (-3,0) node [right] {Input:};
\begin{scope}[shift={(1,0)}]
\path (-0.15,0) node [left, yshift=1pt] {$H=$};
\node[NormalNode, label={}] (0) at (0,0) {};
\node[NormalNode, label={above:$v$}] (1) at (0.5,0) {};
\node[NormalNode, label={}] (2) at (1,0) {};
\node[NormalNode, label={}] (3) at (1.5,0) {};
\draw (0) -- (1) -- (2) -- (3);
\end{scope}
\begin{scope}[shift={(4.5, 0.25)}]
\begin{scope}[shift={(-4pt, -0.25)}]
\draw [decorate,decoration={brace,amplitude=3pt},xshift=0pt]
    (0,-0.35) -- (0,0.35)
    node [black,midway,xshift=-22pt,yshift=0pt] {$F=$};
\draw [decorate,decoration={brace,amplitude=3pt, mirror}, xshift=136pt]
    (0,-0.35) -- (0,0.35);
\end{scope}

\begin{scope}[shift={(0,0)}]
\node[NormalNode, label={}] (0) at (0,0) {};
\node[NormalNode, label={}] (1) at (0,-0.5) {};
\node[NormalNode, label={}] (2) at (0.5,-0.5) {};
\node[NormalNode, label={}] (3) at (0.5,0) {};
\draw (0) -- (1) -- (2);
\draw[] (3) -- (1);
\path (0.75,-0.5) node [] {$,$};

\end{scope}

\begin{scope}[shift={(1,0)}]
\node[NormalNode, label={}] (0) at (0,0) {};
\node[NormalNode, label={}] (1) at (0,-0.5) {};
\node[NormalNode, label={}] (2) at (0.5,-0.5) {};
\node[NormalNode, label={}] (3) at (0.5,0) {};
\draw (0) -- (1) -- (2);
\draw[] (3) -- (2);
\draw[] (3) -- (0);
\path (0.75,-0.5) node [] {$,$};
\end{scope}

\begin{scope}[shift={(2,0)}]
\node[NormalNode, label={}] (0) at (0,0) {};
\node[NormalNode, label={}] (1) at (0,-0.5) {};
\node[NormalNode, label={}] (2) at (0.5,-0.5) {};
\node[NormalNode, label={}] (3) at (0.5,0) {};
\draw (0) -- (1) -- (2);
\draw[] (3) -- (2);
\draw[] (3) -- (1);
\path (0.75,-0.5) node [] {$,$};

\end{scope}

\begin{scope}[shift={(3,0)}]
\node[NormalNode, label={}] (0) at (0,0) {};
\node[NormalNode, label={}] (1) at (0,-0.5) {};
\node[NormalNode, label={}] (2) at (0.5,-0.5) {};
\node[NormalNode, label={}] (3) at (0.5,0) {};
\draw (0) -- (1) -- (2);
\draw[] (3) -- (2);
\draw[] (3) -- (1);
\draw[] (3) -- (0);
\path (0.75,-0.5) node [] {$,$};

\end{scope}

\begin{scope}[shift={(4,0)}]
\node[NormalNode, label={}] (0) at (0,0) {};
\node[NormalNode, label={}] (1) at (0,-0.5) {};
\node[NormalNode, label={}] (2) at (0.5,-0.5) {};
\node[NormalNode, label={}] (3) at (0.5,0) {};
\draw (0) -- (1) -- (2) -- (0);
\draw[] (3) -- (2);
\draw[] (3) -- (1);
\draw[] (3) -- (0);
\end{scope}

\end{scope}

\end{scope}

\begin{scope}[shift={(0,-1.)}]
\path (-3,0) node [right] {1-expansions---$L_v$:};

\begin{scope}[shift={(1,0)}]
\node[NormalNode, label={}] (0) at (0,0) {};
\node[NormalNode, label={above:$v$}] (1) at (0.5,0) {};
\node[NormalNode, label={}] (2) at (1,0) {};
\node[NormalNode, label={}] (3) at (1.5,0) {};
\node[NormalNode, label={below:$w$}] (4) at (0.5,-0.5) {};
\draw (0) -- (1) -- (2) -- (3);
\draw[] (4) -- (1);

\end{scope}

\begin{scope}[shift={(3,0)}]
\node[NormalNode, label={}] (0) at (0,0) {};
\node[NormalNode, label={above:$v$}] (1) at (0.5,0) {};
\node[NormalNode, label={}] (2) at (1,0) {};
\node[NormalNode, label={}] (3) at (1.5,0) {};
\node[NormalNode, label={below:$w$}] (4) at (0.5,-0.5) {};
\draw (0) -- (1) -- (2) -- (3);
\draw[] (4) -- (1);
\draw[] (4) -- (0);

\end{scope}

\begin{scope}[shift={(5,0)}]
\node[NormalNode, label={}] (0) at (0,0) {};
\node[NormalNode, label={above:$v$}] (1) at (0.5,0) {};
\node[NormalNode, label={}] (2) at (1,0) {};
\node[NormalNode, label={}] (3) at (1.5,0) {};
\node[NormalNode, label={below:$w$}] (4) at (0.5,-0.5) {};
\draw (0) -- (1) -- (2) -- (3);
\draw[] (4) -- (1);
\draw[] (4) -- (0);
\draw[] (4) -- (2);

\end{scope}

\begin{scope}[shift={(7,0)}]
\path (0.5,0) node [left, ] {$\ldots$};
\end{scope}

\end{scope}

\begin{scope}[shift={(0,1)}]

\draw [decorate,decoration={brace,amplitude=5pt, mirror}]
    (0.75,-2.95) -- (8,-2.95);
\path (4.25,-3.45) node [] {All of them expansions of some graphs in $F$? $\Rightarrow$ Color $v$ red.};
\end{scope}

\begin{scope}[shift={(0,-3.3)}]
\path (-3,0) node [right] {Output:};
\begin{scope}[shift={(2,0)}]
\path (-0.15,0) node [left, yshift=1pt] {$H$ with $R=$};
\path (1.75,0) node [right, yshift=0.5pt] {after considering all of $V(H)$.};
\node[NormalNode, label={}] (0) at (0,0) {};
\node[RedNode, label={}] (1) at (0.5,0) {};
\node[RedNode, label={}] (2) at (1,0) {};
\node[NormalNode, label={}] (3) at (1.5,0) {};
\draw (0) -- (1) -- (2) -- (3);
\end{scope}
\end{scope}

\end{tikzpicture}
 \caption{Illustration of the \textit{Color} function.}
 \label{fig:color}
\end{figure}

Let $H$ be a connected graph and $F$ be a set of connected graphs.
The function \textit{Color} tries to color (put into set $R$) as many vertices of $H$ as possible, based on structures that are already handled by graphs in $F$.
In particular, if all 1-expansions of $H$, where a vertex $v \in V(H)$ has more neighbors than in $H$, are already also expansions of some graph in $F$
(i.e., each contains at least one induced subgraph isomorphic to some graph in $F$),
then we do not have to consider the situation where $v$ has more neighbors anymore.
In other words, we can safely put $v$ into $R$ (see \autoref{fig:color} for an illustration).
More formally, let $F^* = \bigcup_{H_F \in F} \sigma^*(H_F)$. For every $v \in V(H)$ define the set $L_v = \{H' \mid H' \in \sigma(H), V(H') = V(H) \cup \{w\}, \{v, w\} \in E(H')\}$.
The function $\mathit{Color}(H, F)$ returns the set $R =\{v \in V(H) \mid L_v \setminus F^* = \emptyset\}$.

We use the \textit{Color} function to color vertices of graphs in $\gLbad$ based on graphs in $\gLgood$.
Having more vertices in set $R$ then enables \textit{Generate} to produce a subgraph branching rule with better branching factor and also increases the possibility to reveal that such a situation is already handled by the hand-made rules.

The following lemma roughly shows that, if the graph is then moved to (the current end of) $\gLgood$, then the set $\tau(\gLgood \cup \gLbad)$ with colored versions of the graphs is exhaustive if and only if the set with uncolored versions of the graphs is exhaustive.

\begin{lemma}
\label{color-lemma}
Let $\tau$ be a brancher and let $L=(H_1, H_2, \ldots, H_r)$ be an ordered list of graphs such that $\Llist = \tau(L)$ is exhaustive. Define the set $L_{<i} = \{H_j \mid j<i\}$ and let $R_i = \mathit{Color}(H_i, L_{<i})$. Construct the list $\Llist' = \tau(H_1, R_1) \circ \tau(H_2, R_2) \circ \ldots \circ \tau(H_r, R_r)$. Then the list $\Llist'$ is exhaustive.
\end{lemma}
\begin{proof}
We will proceed by induction from $i=r$ to $i=1$ to show that $\Llist_i = \tau((H_1, H_2, \ldots, H_{i-1})) \circ \tau(H_i, R_i) \circ \tau(H_{i+1}, R_{i+1}) \circ \ldots \circ \tau(H_r, R_r)$ is exhaustive.
Note that $\Llist_1=\Llist'$.
To simplify the proof, we denote $\Llist_{r+1}=\Llist$.
Then the claim holds for $i=r+1$ by assumption, constituting the base case of the induction.

Now, let $1 \leq i \le r$, assume that the claim holds for all strictly greater $i$'s and, for contradiction, that $\Llist_i = \tau((H_1, H_2, \ldots, H_{i-1})) \circ \tau(H_i, R_i) \circ \tau(H_{i+1}, R_{i+1}) \circ \ldots \circ \tau(H_r, R_r)$ is not exhaustive.
All the rules in $\Llist_i$ are correct by definition.
Let $G$ be a connected \bumpy{} graph with at least $\Psi(\Llist_i)=\Psi(\Llist)$ vertices to which none of the rules in $\Llist_i$ applies.
Let $\varrho = (H_j,R_j, \B_j)$ be the first rule in $\Llist_{i+1}$ that applies to~$G$.
If $j \neq i$, then $\varrho \in \Llist_i$ and we get a contradiction.
Therefore, it must be that $j = i$, i.e., we have that $\varrho = \tau(H_i)$ applies to $G$ and $\tau(H_i, R_i)$ does not.

Let $G'$ be the induced subgraph of $G$ to which $\tau(H_i)$ applies and let $\phi$ be the witnessing isomorphism. Since $\tau(H_i, R_i)$ does not apply to $G'$, there must be a vertex $v \in V(G')$ such that $\phi^{-1}(v) \in R_i$ and for which there exists a vertex $u \in V(G)$ such that $u \notin V(G')$ and edge $\{v,u\} \in E(G)$. But as $\phi^{-1}(v) \in R_i$, by the construction of $R_i$, for each supergraph $H'$ in the set $\{H' \mid H' \in \sigma(H_i), V(H') = V(H_i) \cup \{w\}, \{\phi^{-1}(v), w\} \in E(H')\}$ there must be a graph in $L_{<i}$ that is an induced subgraph of $H'$. Consequently, there must be some rule in $\tau(L_{<i})$ which applies to $G[V(G') \cup \{v\}]$, which is a contradiction with the fact that $\varrho$ is the first rule in $\Llist_{i+1}$ that applies to $G$.
\qed
\end{proof}

\subsection{\textit{Generate} function}\label{subs:generate}

The function \textit{Generate} represents our brancher.
For simplification, it also takes care about exploiting the rules in $\A$ (the hand-made rules).

\begin{definition}
Let $H$ be a connected \bumpy{} graph and $R \subseteq V(H)$. Let $(H, R, \B)$ be any correct subgraph branching rule constructed for the pair $H, R$. Let $G$ be an input graph of any instance of given $\FSVD$ problem. The pair $H, R$ is called \emph{handled with respect to $\A$} if whenever the rule $(H, R, \B)$ would apply to $G$, some rule from $\A$ would also apply to $G$.
\end{definition}

\noindent
Let $H$ be a connected \bumpy{} graph and $R \subseteq V(H)$. The function $\mathit{Generate}_{\A}(H, R)$ returns either a correct subgraph branching rule $(H, R, \B)$, or determines that the pair $H,R$ is handled with respect to $\A$. This fact is signaled by returning ``HANDLED''. The way the subgraph branching rules are constructed, i.e., the function $\mathit{Generate}_{\A}$, will be described in Section \ref{generate-function-section}.
For the purpose of the analysis of the algorithm, we also use function $\mathit{Generate}'_{\A}$ which returns the same rule as $\mathit{Generate}_{\A}$ if $\mathit{Generate}_{\A}(H,R) \neq $ HANDLED and the rule from Observation~\ref{correct-subgraph-branching-rule-existence-observation} otherwise. By definition, $\mathit{Generate}'_{\A}$ is a brancher.

\subsection{\textit{Expand} function}\label{subs:expand}

\begin{figure}
 \begin{tikzpicture}

\path (-3,-0.22) node [right] {Input:};
\begin{scope}[shift={(2.5,0)}]
\begin{scope}[shift={(0, -0.25)}]
\draw [decorate,decoration={brace,amplitude=3pt},xshift=0pt]
    (0,-0.35) -- (0,0.35)
    node [black,midway,xshift=-22pt,yshift=1pt] {$L =$};
\path (24pt,-0,2) node [] {$,$};

\draw [decorate,decoration={brace,amplitude=3pt, mirror}, xshift=53pt]
    (0,-0.35) -- (0,0.35);
\end{scope}

\begin{scope}[shift={(4pt, 0)}]
\node[NormalNode, label={}] (0) at (0,0) {};
\node[NormalNode, label={}] (1) at (0,-0.5) {};
\node[NormalNode, label={}] (2) at (0.5,-0.5) {};
\draw (0) -- (1) -- (2);
\end{scope}

\begin{scope}[shift={(34pt, 0)}]
\node[NormalNode, label={}] (0) at (0,0) {};
\node[NormalNode, label={}] (1) at (0,-0.5) {};
\node[NormalNode, label={}] (2) at (0.5,-0.5) {};
\draw (0) -- (1) -- (2) -- (0);
\end{scope}

\end{scope}
\begin{scope}[shift={(8,0)}]
\begin{scope}[shift={(0, -0.25)}]
\draw [decorate,decoration={brace,amplitude=3pt},xshift=0pt]
    (0,-0.35) -- (0,0.35)
    node [black,midway,xshift=-22pt,yshift=1pt] {$F=$};

\draw [decorate,decoration={brace,amplitude=3pt, mirror}, xshift=23pt]
    (0,-0.35) -- (0,0.35);
\end{scope}

\begin{scope}[shift={(4pt, 0)}]
\node[NormalNode, label={}] (0) at (0,0) {};
\node[NormalNode, label={}] (1) at (0,-0.5) {};
\node[NormalNode, label={}] (2) at (0.5,-0.5) {};
\node[NormalNode, label={}] (3) at (0.5, 0) {};
\draw (0) -- (1) -- (2) -- (3) -- (0);
\end{scope}

\end{scope}

\path (-3,-2.7) node [right] {1-expansions:};

\begin{scope}[shift={(0,-2)}]

\begin{scope}[shift={(0,0)}]
\node[NormalNode, label={}] (0) at (0,0) {};
\node[NormalNode, label={}] (1) at (0,-0.5) {};
\node[NormalNode, label={}] (2) at (0.5,-0.5) {};
\node[NormalNode, label={}] (3) at (0.5,0) {};
\draw (0) -- (1) -- (2);
\draw[] (3) -- (2);
\path (0.75,-0.5) node [] {$,$};
\end{scope}

\begin{scope}[shift={(0,-1)}]
\node[NormalNode, label={}] (0) at (0,0) {};
\node[NormalNode, label={}] (1) at (0,-0.5) {};
\node[NormalNode, label={}] (2) at (0.5,-0.5) {};
\node[NormalNode, label={}] (3) at (0.5,0) {};
\draw (0) -- (1) -- (2);
\draw[] (3) -- (1);
\path (0.75,-0.5) node [] {$,$};
\end{scope}

\begin{scope}[shift={(1,0)}]
\node[NormalNode, label={}] (0) at (0,0) {};
\node[NormalNode, label={}] (1) at (0,-0.5) {};
\node[NormalNode, label={}] (2) at (0.5,-0.5) {};
\node[NormalNode, label={}] (3) at (0.5,0) {};
\draw (0) -- (1) -- (2);
\draw[] (3) -- (2);
\draw[] (3) -- (1);
\path (0.75,-0.5) node [] {$,$};

\end{scope}

\begin{scope}[shift={(1,-1)}]
\node[NormalNode, label={}] (0) at (0,0) {};
\node[NormalNode, label={}] (1) at (0,-0.5) {};
\node[NormalNode, label={}] (2) at (0.5,-0.5) {};
\node[NormalNode, label={}] (3) at (0.5,0) {};
\draw (0) -- (1) -- (2);
\draw[] (3) -- (2);
\draw[] (3) -- (0);

\end{scope}

\begin{scope}[shift={(2,0)}]
\node[NormalNode, label={}] (0) at (0,0) {};
\node[NormalNode, label={}] (1) at (0,-0.5) {};
\node[NormalNode, label={}] (2) at (0.5,-0.5) {};
\node[NormalNode, label={}] (3) at (0.5,0) {};
\draw (0) -- (1) -- (2);
\draw[] (3) -- (2);
\draw[] (3) -- (1);
\draw[] (3) -- (0);
\path (0.75,-0.5) node [] {$,$};
\end{scope}

\end{scope}

\tikzset{> = latex}
\draw[->] (2.9,-0.7) .. controls(2.9,-1.3) .. (1.85, -1.3) .. controls(1.25,-1.3) ..  (1.25,-1.8);

\begin{scope}[shift={(4,-2)}]

\begin{scope}[shift={(0,0)}]
\node[NormalNode, label={}] (0) at (0,0) {};
\node[NormalNode, label={}] (1) at (0,-0.5) {};
\node[NormalNode, label={}] (2) at (0.5,-0.5) {};
\node[NormalNode, label={}] (3) at (0.5,0) {};
\draw (0) -- (1) -- (2) -- (0);
\draw[] (3) -- (2);
\path (0.75,-0.5) node [] {$,$};

\end{scope}

\begin{scope}[shift={(1,0)}]
\node[NormalNode, label={}] (0) at (0,0) {};
\node[NormalNode, label={}] (1) at (0,-0.5) {};
\node[NormalNode, label={}] (2) at (0.5,-0.5) {};
\node[NormalNode, label={}] (3) at (0.5,0) {};
\draw (0) -- (1) -- (2) -- (0);
\draw[] (3) -- (2);
\draw[] (3) -- (1);
\path (0.75,-0.5) node [] {$,$};

\end{scope}

\begin{scope}[shift={(2,0)}]
\node[NormalNode, label={}] (0) at (0,0) {};
\node[NormalNode, label={}] (1) at (0,-0.5) {};
\node[NormalNode, label={}] (2) at (0.5,-0.5) {};
\node[NormalNode, label={}] (3) at (0.5,0) {};
\draw (0) -- (1) -- (2) -- (0);
\draw[] (3) -- (2);
\draw[] (3) -- (1);
\draw[] (3) -- (0);
\end{scope}

\end{scope}

\draw[->] (3.9,-0.7) .. controls(3.9,-1.3) .. (4.85, -1.3) .. controls(5.25,-1.3) ..  (5.25,-1.8);

\path (-3,-4.55) node [right] {Non-isomorphic:};

\draw [decorate,decoration={brace,amplitude=5pt, mirror}]
    (-0.1,-3.8) -- (6.6,-3.8);

\begin{scope}[shift={(0.5,-4.3)}]

\begin{scope}[shift={(0,0)}]
\node[NormalNode, label={}] (0) at (0,0) {};
\node[NormalNode, label={}] (1) at (0,-0.5) {};
\node[NormalNode, label={}] (2) at (0.5,-0.5) {};
\node[NormalNode, label={}] (3) at (0.5,0) {};
\draw (0) -- (1) -- (2);
\draw[] (3) -- (2);
\path (0.75,-0.5) node [] {$,$};
\end{scope}

\begin{scope}[shift={(1,0)}]
\node[NormalNode, label={}] (0) at (0,0) {};
\node[NormalNode, label={}] (1) at (0,-0.5) {};
\node[NormalNode, label={}] (2) at (0.5,-0.5) {};
\node[NormalNode, label={}] (3) at (0.5,0) {};
\draw (0) -- (1) -- (2);
\draw[] (3) -- (1);
\path (0.75,-0.5) node [] {$,$};
\end{scope}

\begin{scope}[shift={(2,0)}]
\node[NormalNode, label={}] (0) at (0,0) {};
\node[NormalNode, label={}] (1) at (0,-0.5) {};
\node[NormalNode, label={}] (2) at (0.5,-0.5) {};
\node[NormalNode, label={}] (3) at (0.5,0) {};
\draw (0) -- (1) -- (2);
\draw[] (3) -- (2);
\draw[] (3) -- (1);
\path (0.75,-0.5) node [] {$,$};

\end{scope}

\begin{scope}[shift={(3,0)}]
\node[NormalNode, label={}] (0) at (0,0) {};
\node[NormalNode, label={}] (1) at (0,-0.5) {};
\node[NormalNode, label={}] (2) at (0.5,-0.5) {};
\node[NormalNode, label={}] (3) at (0.5,0) {};
\draw (0) -- (1) -- (2);
\draw[] (3) -- (2);
\draw[] (3) -- (0);
\path (0.75,-0.5) node [] {$,$};

\end{scope}

\begin{scope}[shift={(4,0)}]
\node[NormalNode, label={}] (0) at (0,0) {};
\node[NormalNode, label={}] (1) at (0,-0.5) {};
\node[NormalNode, label={}] (2) at (0.5,-0.5) {};
\node[NormalNode, label={}] (3) at (0.5,0) {};
\draw (0) -- (1) -- (2);
\draw[] (3) -- (2);
\draw[] (3) -- (1);
\draw[] (3) -- (0);
\path (0.75,-0.5) node [] {$,$};
\end{scope}

\begin{scope}[shift={(5,0)}]
\node[NormalNode, label={}] (0) at (0,0) {};
\node[NormalNode, label={}] (1) at (0,-0.5) {};
\node[NormalNode, label={}] (2) at (0.5,-0.5) {};
\node[NormalNode, label={}] (3) at (0.5,0) {};
\draw (0) -- (1) -- (2) -- (0);
\draw[] (3) -- (2);
\draw[] (3) -- (1);
\draw[] (3) -- (0);
\end{scope}

\end{scope}

\path (-3,-5.85) node [right] {Removal of $F^*$:};

\draw[->] (8.4,-0.7) .. controls(8.4,-4.55) .. (6.8, -5.05);

\draw [decorate,decoration={brace,amplitude=5pt, mirror}]
    (-0.1,-5.1) -- (6.6,-5.1);

\begin{scope}[shift={(1.0,-5.6)}]

\begin{scope}[shift={(0,0)}]
\node[NormalNode, label={}] (0) at (0,0) {};
\node[NormalNode, label={}] (1) at (0,-0.5) {};
\node[NormalNode, label={}] (2) at (0.5,-0.5) {};
\node[NormalNode, label={}] (3) at (0.5,0) {};
\draw (0) -- (1) -- (2);
\draw[] (3) -- (2);
\path (0.75,-0.5) node [] {$,$};
\end{scope}

\begin{scope}[shift={(1,0)}]
\node[NormalNode, label={}] (0) at (0,0) {};
\node[NormalNode, label={}] (1) at (0,-0.5) {};
\node[NormalNode, label={}] (2) at (0.5,-0.5) {};
\node[NormalNode, label={}] (3) at (0.5,0) {};
\draw (0) -- (1) -- (2);
\draw[] (3) -- (1);
\path (0.75,-0.5) node [] {$,$};
\end{scope}

\begin{scope}[shift={(2,0)}]
\node[NormalNode, label={}] (0) at (0,0) {};
\node[NormalNode, label={}] (1) at (0,-0.5) {};
\node[NormalNode, label={}] (2) at (0.5,-0.5) {};
\node[NormalNode, label={}] (3) at (0.5,0) {};
\draw (0) -- (1) -- (2);
\draw[] (3) -- (2);
\draw[] (3) -- (1);
\path (0.75,-0.5) node [] {$,$};
\end{scope}

\begin{scope}[shift={(3,0)}]
\node[NormalNode, label={}] (0) at (0,0) {};
\node[NormalNode, label={}] (1) at (0,-0.5) {};
\node[NormalNode, label={}] (2) at (0.5,-0.5) {};
\node[NormalNode, label={}] (3) at (0.5,0) {};
\draw (0) -- (1) -- (2);
\draw[] (3) -- (2);
\draw[] (3) -- (1);
\draw[] (3) -- (0);
\path (0.75,-0.5) node [] {$,$};
\end{scope}

\begin{scope}[shift={(4,0)}]
\node[NormalNode, label={}] (0) at (0,0) {};
\node[NormalNode, label={}] (1) at (0,-0.5) {};
\node[NormalNode, label={}] (2) at (0.5,-0.5) {};
\node[NormalNode, label={}] (3) at (0.5,0) {};
\draw (0) -- (1) -- (2) -- (0);
\draw[] (3) -- (2);
\draw[] (3) -- (1);
\draw[] (3) -- (0);
\end{scope}

\end{scope}

\end{tikzpicture}
 \caption{Illustration of the \textit{Expand} function.}
 \label{fig:expand}
\end{figure}
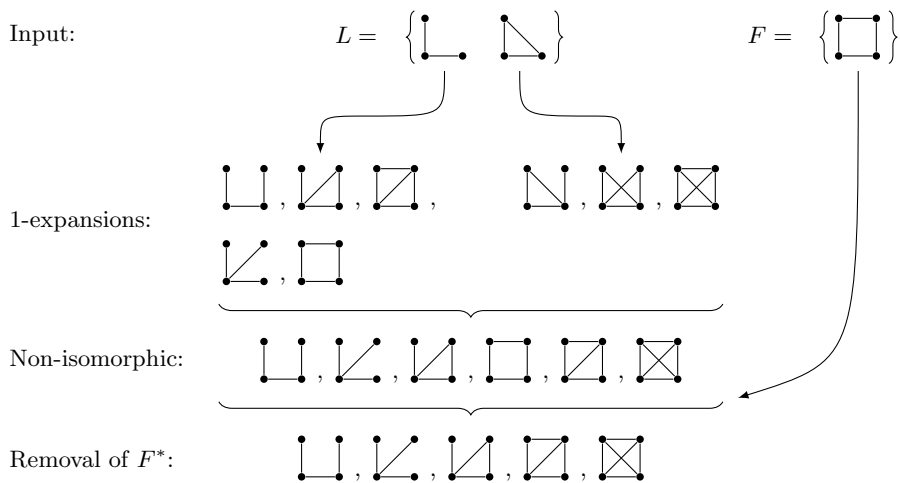

Let $L$ and $F$ be sets of connected graphs.
Function \textit{Expand} returns the set of all 1-expansions of graphs in $L$, except for graphs that are expansions of graphs in $F$ (see \autoref{fig:expand} for an illustration).
More formally, let $F^* = \bigcup_{H_F \in F} \sigma^*(H_F)$. The function $\mathit{Expand}(L, F)$ returns the set of graphs $S = \bigcup_{H_L \in L} \sigma(H_L) \setminus F^*$.
Note that $S$ is obtained from $\bigcup_{H_L \in L} \sigma(H_L)$ by removing all graphs isomorphic to a graph from $F^*$.

We use the function on graphs in $\gLbad$ removing the expansions of graphs in $\gLgood$. The following lemma shows that this does not break the exhaustiveness of the set $\tau(\gLgood \cup \gLbad)$.

\begin{lemma}
\label{expand-lemma}
Let $\tau$ be a brancher and $\gLgood, \gLbad$ be two sets of graphs such that the set of subgraph branching rules $\Llist = \tau(\gLgood \cup \gLbad)$ is exhaustive. Let $\gLbad'$ be the result of calling the function $\mathit{Expand}(\gLbad, \gLgood)$. Then the set of subgraph branching rules $\Llist' = \tau(\gLgood\cup \gLbad')$ is exhaustive.
\end{lemma}
\begin{proof}
Let $G$ be a connected \bumpy{} graph to which no handmade rule in $\A$ applies and which has at least $\Psi(\Llist')$ vertices. For contradiction assume that $\Llist'$ is not exhaustive.
Since by definition all the rules in $\Llist'$ are correct, there must be no rule in $\Llist'$ which applies to $G$.
Let $\varrho$ be a rule in $\Llist$ which applies to $G$, such a rule must exist since $\Llist$ is exhaustive.
If $\varrho \in \Llist'$, then we get a contradiction.
Thus suppose that $\varrho \notin \Llist'$. This means that $\varrho \in \tau(\gLbad)$. Assume that $\varrho = (H,R,\B)$. Let $G'$ be the induced subgraph of $G$ to which $\varrho$ applies. As $\Psi(\tau(\gLbad)) < \Psi(\tau(\gLbad')) \leq \Psi(\Llist')$ and $G$ is connected, there exists a vertex $v \in V(G)$ such that $v \notin V(G')$ and $G[V(G') \cup \{v\}]$ is connected. By construction, $\gLbad'$ contains all the connected supergraphs of $H$ of size $|V(H)|+1$ except those that contain some graph from $\gLgood$ as an induced subgraph. There must be a graph $H' \in \gLbad'$ isomorphic to $G[V(G') \cup \{v\}]$. Suppose that there is not. That would mean that some graph from $\gLgood$ is an induced subgraph of $H'$ and consequently some rule from $\tau(\gLgood)$ would apply to $G[V(G') \cup \{v\}]$, which is a contradiction with no rule in $\Llist'$ applying to $G$. As $H'$ is isomorphic to $G[V(G') \cup \{v\}]$, the rule $\tau(H')$ applies to $G$. And since $\tau(H') \in \tau(\gLbad')$ and therefore $\tau(H') \in \Llist'$, we again arrive at a contradiction with the fact that no rule in $\Llist'$ applies to~$G$.
\qed
\end{proof}

\subsection{Pseudocode and Correctness of the Algorithm}

\begin{algorithm}[t!]
\caption{$(\F,\A,\beta)$-Algorithm}\label{fab-algorithm-pseudocode}
\begin{algorithmic}[1]
  \State $i \gets 0$; $\Llist \gets \emptyset$; $\Llist' \gets \emptyset$; $\gLgood \gets \emptyset$
  \State $\gLbad \gets $ the set $L$ from Observation \ref{initial-list-observation} for $\F$.
  \While{$\gLbad \neq \emptyset$}
    \For{$H \in \gLbad$} \label{main-loop-fab-algorithm-pseudocode}
        \State $R = \mathit{Color}(H, \gLgood)$
        \State $\varrho = \mathit{Generate}_{\A}(H, R)$
        \If{$\bff(\varrho) \leq \beta$ or $\varrho$ = HANDLED}
            \State $i \gets i + 1$
            \State $H_i = H$, $\varrho_i = \varrho$,
            \State $\varrho_i' = \mathit{Generate}'_{\A}(H_i, R)$,
            \State $\gLgood \gets \gLgood \circ (H_i)$ \label{ok1-line-fab-algorithm-pseudocode}
            \State $\gLbad \gets \gLbad \setminus H_i$ \label{removal-to-good}
            \State $\Llist' \gets \Llist' \circ \varrho_i'$
            \If{$\varrho \neq$ HANDLED}
              \State $\Llist \gets \Llist \circ \varrho_i$
            \EndIf \label{ok2-line-fab-algorithm-pseudocode}
            \State \textbf{restart the forcycle}
        \EndIf
    \EndFor
    \State $\gLbad \gets \mathit{Expand}(\gLbad, \gLgood)$ \label{expand-line-fab-algorithm-pseudocode}
  \EndWhile
  \State \Return $\Llist$.
\end{algorithmic}
\end{algorithm}

\noindent
Now we are ready to provide the pseudocode of the $(\F,\A,\beta)$-Algorithm in Algorithm~\ref{fab-algorithm-pseudocode} and prove its correctness.
The algorithm also creates lists $\Llist'$ (only needed for analysis purpose) and $\Llist$ containing the actual subgraph branching rules corresponding to graphs in $\gLgood$ and those not overridden by the rules in $\A$ (by the hand-made rules), respectively.

\begin{theorem}\label{thm:exhaustive}
If the call to $(\F,\A,\beta)$-Algorithm finishes its computation, then the returned ordered list $\Llist$ of subgraph branching rules is exhaustive with respect to $\A$ and for every $\varrho \in \Llist$ we have $\bff(\varrho) \leq \beta$.
\end{theorem}
 \begin{proof}
We start by proving the following statement. Let $\tau$ be a brancher. Then the set $\tau(\gLgood \cup \gLbad)$ is exhaustive at every step of the computation. The proof will be done by induction on the number of steps of the computation. Firstly, at the start of the computation the statement holds due to Observation \ref{initial-list-observation}. Next, observe, that the only place where the set changes during the computation is on line \ref{expand-line-fab-algorithm-pseudocode} where the $\mathit{Expand}$ function is called. By induction the statement holds before the expansion takes place and due to Lemma \ref{expand-lemma} the statement holds after the expansion takes place. Therefore the statement holds at every step of the computation.

Now focus on the contents of $\gLgood, \gLbad$, and $\Llist'$ at the end of the computation.
Let $r$ be the value of $i$ at the end of the computation.
We have that $\gLgood = (H_1, H_2, \ldots, H_r), \gLbad = \emptyset$.
As we have established above, since $\mathit{Generate}'_{\A}$ is a brancher, the set $\mathit{Generate}'_{\A}(\gLgood\cup \gLbad) = \mathit{Generate}'_{\A}(\gLgood)$ is exhaustive. Observe, that for each $\varrho_j = (H_j, R_j, \B_j) \in \Llist', j \in \{1,2, \ldots, r\}$ we have that $R_j$ is exactly $\mathit{Color}(H_j, \{H_k \mid k < j\})$.
Indeed, at that moment when the rule $\varrho'_j$ is added to the list $\Llist'$, we have $\gLgood = (H_1, H_2, \ldots, H_{j-1})$.
This means that at the end of the computation, we have $\gLgood = (H_1, H_2, \ldots, H_r)$ and $\Llist' = \mathit{Generate}'_{\A}(H_1, R_1) \circ \mathit{Generate}'_{\A}(H_2, R_2) \circ \ldots \circ \mathit{Generate}'_{\A}(H_r, R_r)$.
As $\mathit{Generate}'_{\A}(\gLgood)$ is exhaustive, by Lemma \ref{color-lemma}, $\Llist'$ is exhaustive.

Finally, it suffices to observe that $\Llist$ is missing exactly the rules from $\Llist'$ that were handled by some rule in $\A$, which means that whenever such rule could be applied to some graph $G$, some handmade rule in $\A$ could also be applied to $G$. And therefore $\Llist$ is exhaustive with respect to $\A$.

The branching factor of the rules follows directly from the computation of the algorithm.
\qed
\end{proof}

\section{Generating subgraph branching rules}\label{generate-function-section}

This section deals with the task of designing a brancher that achieves branching factors close to the best possible.
Recall that we call our brancher $\mathit{Generate}_{\A}$.

The input of the $\mathit{Generate}_{\A}$ function is the graph $H$ together with its red vertices $R \subseteq V(H)$.
The $\mathit{Generate}_{\A}$ function aims to create a correct subgraph branching rule for the given pair $H, R$ with the smallest possible branching factor.
In general, on one hand, the more branches, the worse the branching factor and on the other hand also the smaller the branches (with less vertices added to the solution) the worse the branching factor.
Ideally, we would like to produce few branches, each containing a lot of vertices.
The task of finding the set of branches with the optimal branching factor is nontrivial and we are not aware of any effective algorithm to compute such a set.
Thus instead of computing the optimal subgraph branching rule, we use heuristics which provide branching rules with reasonable branching factors with affordable computational effort.

The only information we are able to exploit in creating the branches is that a global solution is also a local solution, which must contain at least one vertex of $H$ as $H$ is \bumpy.
Recall that the definition of correctness of a subgraph branching rule requires that for every global solution there is a (global) solution (possibly the same one) and a branch which is a subset of the latter solution.
The ability to move to a different more suitable global solution allows us to reduce the number of necessary branches.

The process of construction of a single subgraph branching rule is split between four functions: $\mathit{Handled}_{\A}$, \textit{Minimal}, \textit{DominanceFree}, and \textit{Adjusted}. The $\mathit{Handled}_{\A}$ detects if the situation is already handled by a handmade rule in $\A$, the function \textit{Minimal} provides an initial set of branches which is then subsequently optimized in the remaining two functions.

\subsection{$\mathit{Handled}_{\A}$ function}\label{subs:handled}
The purpose of this function is to determine whether the pair $H, R$ is already handled by some handmade rule in $\A$.
Since the application of handmade rules in~$\A$ takes precedence before any subgraph branching rules in the $(\A,\Llist)$-Algorithm for \textsc{$\F$-SVD}, this means that before any application of the rule $(H, R, \B)$ would take place, some other handmade rule from $\A$ would be applied instead. This results in $(H, R, \B)$ never being applied and thus there is no need to include any subgraph branching rule for the pair $H, R$ in the resulting set. 
The function returns \textit{YES} if the pair $H, R$ is handled and \textit{NO} otherwise.

Note that this function is a part of the input to the algorithm, representing the set $\A$.

\subsection{\textit{Minimal} function}\label{subs:minimal}

If the pair $(H,R)$ is not handled, then we will try to create a set of branches for it.
We could start with all local solutions.
However, there is no need to include a set in $\B$ if its subset is already included.
Hence the function \textit{Minimal} starts the process with all minimal local solutions.

The input of this function is a connected \bumpy{} graph $H$ and $R \subseteq V(H)$. Let $\B^* = \{S \mid S \subseteq V(H) \land S \text{ is a solution for } H\}$. Now construct the branches $\B_{\mathit{min}}$ as the set of all minimal elements of $\B^*$ according to the $\subseteq$ relation on $\B^*$.
That is, a set $B \in \B^*$ is included in $\B_{\mathit{min}}$ if there is no set $B' \in \B^*$ with $B' \subsetneq B$.
The \textit{Minimal} function returns the branches $\B_{\mathit{min}}$.
\begin{lemma}
\label{minimal-solutions-correct-lemma}
The subgraph branching rule $(H,R,\B_{\mathit{min}})$ is correct.
\end{lemma}
\begin{proof}
Let $G$ be a graph such that $\varrho = (H,R,\B_{\mathit{min}})$ applies to $G$ and let $S$ be the solution for $G$. Let $G' = G[\phi(H)]$ be the induced subgraph to which $\varrho$ applies and $\phi$ be the witnessing isomorphism. Since $H$ is \bumpy{}, $G'$ is also \bumpy{} and $S \cap V(G') \neq \emptyset$. Let $S_H = \phi^{-1}(S \cap V(G'))$.
Observe that $S_H$ must be a solution for $H$ and, hence, is in $\B^*$.
Therefore one of the branches $B \in \B_{\mathit{min}}$ must be a subset of $S_H$. But then $\phi(B) \subseteq \phi(S_H) = S \cap V(G') \subseteq S$. Thus the rule $\varrho$ is correct.
\qed
\end{proof}

\subsection{\textit{DominanceFree} function}\label{subs:dominance}
The input of the function is a connected \bumpy{} graph $H$, $R \subseteq V(H)$, and $\B$ such that $(H,R,\B)$ is a correct subgraph branching rule.
The \textit{DominanceFree} function exploits the ability to (locally) move to a different more suitable global solution in order to reduce the number of necessary branches.
This is also the place where we make use of the red vertices.
In particular, if a vertex $v$ has no neighbors outside $H$, then it might be more beneficial to have a different vertex in the solution, which breaks all copies of graphs from $\F$ as $v$ does and possibly some more partially outside $H$.

This is captured by the dominance between branches.
The basic idea is to take a subset $R^*$ of the red vertices and replace all vertices of the solution in this set by the open neighborhood $N_H(R^*) \setminus R^*$.
We only want to do that if this does not increase the size of the solution and $H[R^*]$ is not \bumpy.
To increase the power of the notion, we do this in a graph $H' = H \setminus B^{\mathit{del}}$, where $B^{\mathit{del}}$ is a set of vertices shared by both the branches (and therefore irrelevant in the moment).

\begin{definition}[Dominated branch]\label{def:dominated_branch}
Let $(H,R,\B)$ be a correct subgraph branching rule. We say that branch $B \in \B$ is \emph{dominated} by branch $B_d \in \B$ if $B_d \neq B$ and there exists a subset $B^{\mathit{del}} \subsetneq B$ such that for $H' = H \setminus B^{\mathit{del}}$, $R' = R \setminus B^{\mathit{del}}$ there exists a subset $R^* \subseteq R', R^* \neq \emptyset$ such that the following holds:
\begin{enumerate}
\item $H[R^*]$ is not \bumpy,
\item $|R^* \cap B| \geq |N_{H'}(R^*) \setminus R^*| \geq 1$,
\item $B_d \subseteq (B \cup N_{H'}(R^*)) \setminus R^*$.
\end{enumerate}
\end{definition}

\noindent Note that if $N_{H'}(R^*) \setminus R^* = \emptyset$, then $B_d \subseteq B$, a case handled by the previous function.

\begin{lemma}
\label{dominance-lemma}
If $(H,R,\B)$ is a correct subgraph branching rule and branch $B \in \B$ is dominated by branch $B_d \in \B$, then $(H,R,\B \setminus \{B\})$ is a correct subgraph branching rule.
\end{lemma}
\begin{proof}
For contradiction suppose that $(H,R,\B \setminus \{B\})$ is not correct.
As $(H,R,\B)$ is correct, there must be a graph $G$ to which $(H,R,\B)$ applies and a solution $S$ for $G$ for which only for branch $B$ there exists a solution $S', |S'| \leq |S|$ such that $\phi(B) \subseteq S'$ where $\phi$ is the witnessing isomorphism. Let $G'$ be the induced subgraph of $G$ such that $G' = \phi(H)$ and we have that $N_G(\phi(R)) \subseteq V(G')$.

Let $B^{\mathit{del}}$ and $R^*$ be the appropriate sets that satisfy the properties for domination of $B$ by $B_d$ and let $H' = H \setminus B^{\mathit{del}}$, $R' = R \setminus B^{\mathit{del}}$.
We are going to show that $S_d = (S' \setminus \phi(B)) \cup  \phi((B \cup N_{H'}(R^*)) \setminus R^*) = (S' \setminus \phi(B \cap R^*)) \cup \phi(N_{H'}(R^*) \setminus R^*)$ is also a solution for $G$ and $|S_d| \leq |S|$.
As $\phi(B_d) \subseteq \phi((B \cup N_{H'}(R^*)) \setminus R^*)$, this will contradict our choice of~$S$.

Suppose that $S_d$ is not a solution for $G$. Then $G \setminus S_d$ is \bumpy{} and $G \setminus S_d$ contains some graph from the \bumpset{} $\F$ as a subgraph, let that subgraph be $F$. There must be a vertex $v \in V(F)$ such that $v \in \phi(B \cap R^*)$, otherwise $S_d$ would be a solution for $G$. Moreover, there must also exist a vertex $u \in V(F)$ such that $u \in V(F) \setminus \phi(R^*)$, otherwise we have $V(F) \subseteq \phi(R^*)$, but $G[\phi(R^*)]$ is not \bumpy{} as $H[R^*]$ is not \bumpy{}. As $v \in \phi(R^*$), $u \notin \phi(R^*)$, and $F$ is connected, we have $F \cap (N_G(\phi(R^*)) \setminus \phi(R^*)) \neq \emptyset$. Since $N_G(\phi(R^*)) \subseteq V(G')$ we have $N_G(\phi(R^*)) \setminus \phi(R^*) = \phi(N_H(R^*) \setminus R^*)$. Finally, we have that $N_H(R^*) \setminus R^* \subseteq B^{\mathit{del}} \cup (N_{H'}(R^*) \setminus R^*)$ and we conclude that $F \cap \phi((B^{\mathit{del}} \cup N_{H'}(R^*)) \setminus R^*) \neq \emptyset$. Thus $F \cap S_d \neq \emptyset$ and we arrive at a contradiction with $F$ being a subgraph of $G \setminus S_d$.

Since $S_d = (S' \setminus \phi(B)) \cup  \phi((B \cup N_{H'}(R^*)) \setminus R^*) = (S' \setminus \phi(B \cap R^*)) \cup \phi(N_{H'}(R^*) \setminus R^*)$ and $|R^* \cap B| \geq |N_{H'}(R^*) \setminus R^*|$ we have that $|S_d| \leq |S|$, concluding the proof.
\qed
\end{proof}

The purpose of the \textit{DominanceFree} function is to remove branches that are dominated by other branches.
However, as there might be cycles of dominance, we have to be a little bit more careful.
Consider directed graph $G_{\B} = (\B, E_{\B})$ such that $(B_i, B_j) \in E_{\B}$ if and only if $B_i$ is dominated by $B_j$. Let $C_1, C_2, \ldots, C_c$ be the strongly connected components of $G_{\B}$. By $\mathit{rep}(C_i)$ we denote an arbitrary, but fixed, branch $B \in C_i$. A component $C_i$ is called a \emph{sink} component if there is no other component $C_j, i \neq j$ such that there exists an edge $(B_i, B_j) \in E_{\B}$ where $B_i \in C_i$ and $B_j \in C_j$.
The \textit{DominanceFree} function returns the branches $\B_{\mathit{df}} = \{rep(C_i) \mid i \in \{1,2,\ldots,c\} \wedge C_i \;{\textup{is a sink component}}\}$.

\begin{lemma}
\label{dominance-solutions-correct-lemma}
The subgraph branching rule $(H, R, \B_{\mathit{df}})$ is correct.
\end{lemma}
\begin{proof}
We will construct a linear ordering of the branches $B \in (\B \setminus \B_{\mathit{df}})$ such that by repeatedly applying Lemma \ref{dominance-lemma} alongside this ordering we arrive at a sequence of correct subgraph branching rules which starts with $(H, R, \B)$ and ends with $(H, R, \B_{\mathit{df}})$.

Observe that for each $B \in (\B \setminus \B_{\mathit{df}})$ there exists an oriented path $(B_1, B_2, \ldots, B_p)$ in $G_{\B}$ such that $B_1 = B$ and $B_p \in \B_{\mathit{df}}$. Let $\delta(B)$ be the shortest distance of $B$ to some vertex in $\B_{\mathit{df}}$ and let $\Delta = \max_{B \in (\B \setminus \B_{\mathit{df}})}\delta(B)$. The ordering we are looking for is then $(B^{\Delta}_1, \ldots, B^{\Delta}_{d_\Delta}, B^{\Delta-1}_{1}, \ldots, B^{\Delta-1}_{d_{\Delta-1}},\ldots, B^{1}_{1}, \ldots, B^{1}_{d_{1}})$ where for $i \in \{\Delta, \Delta-1, \ldots, 1\}, j \in \{1,2,\ldots, d_{i}\}, \delta({B^i_j}) = i$.

To prove the fact that each rule in the resulting sequence of subgraph branching rules is correct, suppose that we want to eliminate branch $B^i_j$. Take a look at the oriented path $(B_1, B_2, \ldots, B_p)$ where $B_1 = B^i_j$ and $B_p \in \B_{\mathit{df}}$. As $\delta(B^i_j) = i$ we have that $B_2 \in \{B^{i-1}_k \mid k \in \{1,2, \ldots, d_{i-1}\}\}$ which means that $B_2$ has not been eliminated yet. Therefore $B_1$ is dominated by $B_2$ and Lemma \ref{dominance-lemma} applies.
\qed
\end{proof}

\subsection{\textit{Adjusted} function}\label{subs:adjust}
So far we were trying to use branches as large possible (exploiting that they must be a local solution) and only tried to reduce their number.
However, by replacing several larger branches with a branch which is their intersection we can sometimes improve the branching factor of the branching rule.
This might, e.g., correspond to the branching rule only ``revealing the solution'' on a suitable part of graph $H$ and ignoring the not so favorably structured rest of $H$.

This is the task of the \textit{Adjusted} function.
The input of this function is a connected \bumpy{} graph $H$, $R \subseteq V(H)$, and $\B$ such that $(H,R,\B)$ is a correct subgraph branching rule.
The function heuristically searches for potential replacement branches as described in Algorithm \ref{fab-adjusted-function-pseudocode}.

\begin{lemma}
\label{adjusted-lemma}
Let $(H,R,\B)$ be a correct subgraph branching rule. For any $A \subseteq V(H), A \neq \emptyset$ construct the branches $\B_A = \{B \mid B \in \B \wedge A \not \subseteq B\} \cup \{A\}$. The subgraph branching rule $(H, R, \B_A)$ is correct.
\end{lemma}
\begin{proof}
For contradiction suppose that $(H, R, \B_A)$ is not correct.
Let $G$ be a graph to which $(H, R, \B_A)$ applies, $\phi$ the witnessing isomorphism, and $S$ a solution for $G$ such that there is no branch $B \in \B_A$ and a solution $S', |S'| \leq |S|$ such that $\phi(B) \subseteq S'$.
As $(H, R, \B)$ is correct, let $B \in \B$ be the branch for which there exists a solution $S', |S'| \leq |S|$ such that $\phi(B) \subseteq S'$.
If $B \in \B_A$, then we get a contradiction, so suppose that $B \notin \B_A$. By construction of $\B_A$ this means that $A \subseteq B$ which immediately gives us that $\phi(A) \subseteq S'$ which is again a contradiction.
\qed
\end{proof}
\noindent
Now consider the pseudocode of the \textit{Adjusted} function in Algorithm \ref{fab-adjusted-function-pseudocode}.
It tries all possible replacement sets $A$ and takes one which improves the branching factor the most.
This is repeated as long as the branching factor is improved.
The \textit{Adjusted} function returns the branches $\B_{\mathit{adj}}$.

\begin{algorithm}[ht]
\caption{\textit{Adjusted} function}\label{fab-adjusted-function-pseudocode}
\begin{algorithmic}[1]
 \Function{$\mathit{Adjusted}$}{$H,R,\B$}
 \State $\B_{\mathit{adj}} \gets \B$
 \State $\mathcal{A} \gets \bigcup_{B\in\B_{\mathit{adj}}} 2^{B} \setminus \{\emptyset\}$ \label{fab-adjusted-function-pseudocode-main-loop}
 \State Find $A \in \mathcal{A}$ that minimizes $\bff((H, R, \B_A))$, where $\B_A = \{B \mid B \in \B \wedge A \not \subseteq B\} \cup \{A\}$.\label{fab-adjusted-function-pseudocode-lemma-point}%
 \If{$\bff((H, R, \B_A)) < \bff((H, R, \B_{\mathit{adj}}))$}
    \State $\B_{\mathit{adj}} \gets \B_A$
    \State \textbf{go to \ref{fab-adjusted-function-pseudocode-main-loop}}
 \EndIf
 \State \Return $\B_{\mathit{adj}}$
 \EndFunction
\end{algorithmic}
\end{algorithm}

\begin{lemma}
\label{adjusted-solutions-correct-lemma}
Let $(H,R,\B)$ be a correct subgraph branching rule. Let $\B_{\mathit{adj}}$ be the result of $\mathit{Adjusted}(H,R,\B)$. The subgraph branching rule $(H, R, \B_{\mathit{adj}})$ is correct.
\end{lemma}
\begin{proof}
The construction of branches $\B_A$ on line \ref{fab-adjusted-function-pseudocode-lemma-point} exactly follows the construction in Lemma \ref{adjusted-lemma} and therefore the resulting subgraph branching rule $(H,R,\B_A)$ is correct. Since this is the only place in the algorithm where the branches change, the returned rule must be correct.
\qed
\end{proof}

\subsection{Complete \textit{Generate} function}\label{subs:complete}
The order of calls to the four functions is described in Algorithm \ref{fab-generate-function-pseudocode}.

\begin{algorithm}[ht]
\caption{$\mathit{Generate}_{\A}$ function}\label{fab-generate-function-pseudocode}
\begin{algorithmic}[1]
 \State Let $H$ be a connected \bumpy{} graph and $R \subseteq V(H)$.
 \Function{$\mathit{Generate}_{\A}$}{$H,R$}
 \If{$\mathit{Handled}_{\A}(H,R)$ = YES}
   \State Return HANDLED.
 \EndIf
 \State $\B_{\mathit{min}} \gets \mathit{Minimal}(H, R)$
 \State $\B_{\mathit{df}} \gets \mathit{DominanceFree}(H, R, \B_{\mathit{min}})$
 \State $\B_{\mathit{adj}} \gets \mathit{Adjusted}(H, R, \B_{\mathit{df}})$
 \State \Return $(H, R, \B_{\mathit{adj}})$
 \EndFunction
\end{algorithmic}
\end{algorithm}

\begin{theorem}\label{thm:output_rule_correct}
If the \textit{Generate} function returns a subgraph branching rule $(H, R, \B)$, the subgraph branching rule $(H, R, \B)$ is correct.
\end{theorem}

\begin{proof}
The statement immediately follows from Lemmata \ref{minimal-solutions-correct-lemma}, \ref{dominance-solutions-correct-lemma}, and \ref{adjusted-solutions-correct-lemma}.
\qed
\end{proof}

\section{Applying $(\F,\A,\beta)$-Algorithm to \textsc{$d$-PVC}}\label{sec:applying}
We are now going to show the specifics of applying the $(\F,\A,\beta)$-Algorithm to the \textsc{$d$-Path Vertex Cover} problem. It is easy to see that \textsc{$d$-PVC} equals \FSVD{} for $\F = \{P_d\}$.

\lv{
A~\textit{$d$-path}, denoted as an ordered~$d$-tuple $P_d = (p_1, p_2, \ldots, p_d)$, is a~path on $d$~vertices $\{p_1, p_2, \ldots, p_d\}$. A~\textit{$P_d$-free} graph is a~graph that does not contain a~$P_d$ as a~subgraph. The \textsc{$d$-Path Vertex Cover} problem is formally defined as follows:

\vspace{2mm}
\noindent
\begin{tabularx}{\textwidth}{|l|X|}
  \hline
\multicolumn{2}{|l|}{\textsc{$d$-Path Vertex Cover, $d$-PVC}} \\ \hline
  \textsc{Input}: & A~graph $G=(V,E)$, an integer $k \in \Z^{+}_0$. \\
  \textsc{Output}: & A~set $S \subseteq V$, such that $|S| \leq k$ and $G \setminus S$ is a~$P_d$-free graph. \\
  \hline
\end{tabularx}
\vspace{2mm}
}

\subsection{Handmade Rules}

 For the $(\F,\A,\beta)$-Algorithm to work for interesting values of $\beta$, we provide two handmade polynomial time reduction rules to $\A$ that are correct for \textsc{$d$-PVC}. After we present the rules, we also discuss the appropriate $\mathit{Handled}_{\A}$ function.

\begin{rrule}[Red component reduction for \textsc{$d$-PVC}]\label{rul:red_component}
Let $(G,k)$ be an instance of \textsc{$d$-PVC}. Let $v \in V(G)$ be a vertex such that there are at least two $P_d$-free connected components $C_1, C_2$ in $G \setminus v$. If there is a $P_d$ in $G[\{v\} \cup V(C_1) \cup V(C_2)]$, reduce $(G,k)$ to instance $(G \setminus (\{v\} \cup V(C_1) \cup V(C_2)), k-1)$ which corresponds to taking $v$ into a solution. Otherwise, let $P^1_i$ be the longest path in $G[\{v\} \cup V(C_1)]$ starting in $v$ and let $P^2_j$ be the longest path in $G[\{v\} \cup V(C_2)]$ starting in $v$. Assume, without loss of generality, that $i \leq j$. Then, reduce the instance $(G, k)$ to $(G \setminus V(C_1), k)$.
\end{rrule}

\begin{proof}[of correctness] 
Set $A_v = \{v\} \cup V(C_1) \cup V(C_2)$. Observe, that any $P_d$ in $G$ that uses some of the vertices in $A_v$ must include $v$ as the components $C_1$ and $C_2$ are $P_d$-free.

For the first case, we need to show that $(G,k)$ has a solution if and only if $(G \setminus A_v, k-1)$ has a solution.

Let $S$ be a solution for $G$ of size at most $k$.
As there is a $P_d$ in $G[A_v]$, at least one of the vertices in $A_v$ must be in $S$. Then, $S' = S \setminus A_v$ is a solution for $G \setminus A_v$ of size at most $k-1$ as $(G \setminus A_v) \setminus S'$ is a subgraph of $G \setminus S$.

Conversely, if $S'$ is a solution for $G \setminus A_v$ of size at most $k-1$, then $S = S' \cup \{v\}$ is a solution for $G$ of size at most $k$ as, by the above observation, any $P_d$ in $G \setminus S'$ must include $v$.

For the second case, we need to show that $(G,k)$ has a solution if and only if $(G \setminus V(C_1), k)$ has a solution.

Let $S$ be a solution for $G$ of size at most $k$. Then $S \setminus V(C_1)$ is a solution for $G \setminus V(C_1)$ as it is a subgraph of $G$.

For the other direction, let $S'$ be a solution for $G \setminus V(C_1)$ of size at most $k$. Now, assume that $S'$ is not a solution for $G$ and there is a $d$-path $P$ in $G \setminus S'$. The path $P$ must be of the following form: $(x_1, x_2, \ldots, x_p, v, c_1, c_2, \ldots, c_q)$ where $\{x_1, x_2, \ldots, x_p\} \subseteq (V(G) \setminus A_v)$ and $\{c_1, c_2, \ldots, c_q\} \subseteq V(C_1)$.

Firstly, assume that $S' \cap V(C_2) = \emptyset$. Recall that we assumed that there is a path $P^2_j$ in $G[\{v\} \cup V(C_2)]$ starting in $v$ which is longer than any path in $G[\{v\} \cup V(C_1)]$ starting in $v$, in particular, longer than $(v, c_1, c_2, \ldots, c_q)$. Let $P^2_j = (v, y_1, y_2, \ldots, y_{j-1})$. But then a $d$-path $P' = (x_1, x_2, \ldots, x_p, v, y_1, y_2, \ldots, y_{d-p-1})$ exists in $(G \setminus V(C_1))\setminus S'$ which is a contradiction with $S'$ being a solution for $G \setminus V(C_1)$.

Lastly, we have that $S' \cap V(C_2) \neq \emptyset$. In this case, by the above observation, $S'' = (S' \setminus V(C_2)) \cup \{v\}$ is also a solution for $G \setminus V(C_1)$ and of size at most $k$. But then $S'' \cap P \neq \emptyset$ and therefore $S''$ is a solution for $G$ of size at most $k$.
\qed
\end{proof}
\begin{rrule}[Red star reduction for \textsc{$d$-PVC}, $d\geq4$]\label{rul:red_star}
Let $(G, k)$ be the instance of \textsc{$d$-PVC}. Suppose there exists a subset $C \subseteq V(G)$, $|C| \leq \dofloor*{\frac{d}{2}}-1$ for which there is a subset $L \subseteq V(G)$ such that $\forall v \in L, N(v) = C$ and $|L| \geq 2|C|$. Let $x \in L$. Then reduce instance $(G, k)$ to instance $(G \setminus \{x\}, k)$.
\end{rrule}

\begin{proof}[of correctness]
We need to show that $(G,k)$ has a solution if and only if $(G \setminus \{x\},k)$ has a solution. Consider the direction from left to right. If $S$ is a solution for $G$, then $S \setminus \{x\}$ is a solution for $G \setminus \{x\}$ since $(G \setminus \{x\}) \setminus (S \setminus \{x\}) = G \setminus S$.

For the other direction, let $S'$ be a solution for $G' = G \setminus \{x\}, |S'| \leq k$. Suppose that $S'$ is not a solution for $G$. Then in $G \setminus S'$ there must be a path $P_d = (v_1, v_2, \ldots, v_d)$ with $v_i = x$.
We show that $|P_d \cap L| \leq |C|$.
Indeed, for $P_d$ to use more than $|C|$ vertices in $L$, it would have to start and end in $G[L]$ and alternate between vertices from $C$ and $L$. Which means that $P_d$ would be contained in $G[L \cup C]$. But as the length $l$ of the longest path possible in $G[L \cup C]$ is $2|C| + 1$ and $|C| \leq \dofloor*{\frac{d}{2}}-1$, we have that $l \leq 2\dofloor*{\frac{d}{2}} -1 \leq d-1$, which contradicts the fact that $P_d$ contains $d$ vertices. As $|P_d \cap L| \leq |C|$ and $2|C| \leq |L|$, we have that $|L \setminus P_d|\geq |C|$. Let $y \in L \setminus P_d$. Suppose that $y \notin S'$. But then in $G' \setminus S'$ there would be a path $P'_d = (v'_1, v'_2, \ldots, v'_d)$ such that $v'_j = v_j$ for $j \neq i$ and $v'_i = y$, which is a contradiction with $S'$ being a solution for $G'$. Therefore we have that $L \setminus P_d \subseteq S'$. Finally, observe that any $d$-path that uses some vertex from $L$ must also use a vertex from $C$ and thus $S = (S' \setminus (L \setminus P_d)) \cup C$ is a solution for $G$ and $|S| \leq |S'|$ as $|L \setminus P_d| \geq |C|$.
\qed
\end{proof}


\noindent
Now that we have established that our reduction rules are correct for \textsc{$d$-PVC}, we have to discuss their operation in the $(\F,\A,\beta)$-Algorithm.
Note that as a part of $\A$, if the rule applies, we would make a call of \textit{SolveRecursively}($G \setminus (\{v\} \cup V(C_1) \cup V(C_2)),k - 1$), \textit{SolveRecursively}($G \setminus V(C_1), k$), or \textit{SolveRecursively}($G \setminus \{x\},k$), respectively, and return the answer obtained. The following two lemmata describe the function $\mathit{Handled}_{\A}$.

\begin{lemma}
\label{lem:red_component_reduction}
For the case of the \textit{red component reduction} rule, let $H$ be a connected \bumpy{} graph and $R \subseteq V(H)$ be its red vertices. If there is a vertex $v \in V(H)$ for which there are at least two $d$-path free connected components $C_1, C_2$ in $H \setminus v$ with  $V(C_1), V(C_2) \subseteq R$ then the pair $H, R$ is handled by the red component reduction rule, i.e., whenever any subgraph branching rule $(H, R, \B)$ would apply to a graph $G$, the red component reduction rule would also apply to $G$.
\end{lemma}
\begin{proof}
Let $G$ be a connected \bumpy{} graph and let $\varrho = (H, R, \B)$ be any rule constructed for the pair $H, R$. Suppose that $(H, R, \B)$ applies to $G$. Let $\phi$ be the witnessing isomorphism.

Focus on the components $C_1, C_2$. As $V(C_1), V(C_2) \subseteq R$ and they are connected components of $H \setminus v$, we have that $N_G(\phi(C_1)) = \{\phi(v)\}$ and $N_G(\phi(C_2)) = \{\phi(v)\}$. That means that $\phi(C_1), \phi(C_2)$ are $d$-path free connected components of $G \setminus \phi(v)$ and the conditions of the \textit{red component reduction} rule are satisfied and therefore the \textit{red component reduction} rule applies to $G$.
\qed
\end{proof}

\begin{lemma}
\label{lem:red_star_handled}
For the case of \textsc{$d$-PVC}, $d \geq 4$. Let $H$ be a connected \bumpy{} graph and $R \subseteq V(H)$ be its red vertices. If there is a subset $C \subseteq V(H), |C| \leq \dofloor*{\frac{d}{2}}-1$ for which there is a subset $L \subseteq R$ such that $\forall v \in L, N(v) = C$ and $2|C| \leq |L|$, then the pair $H, R$ is handled by the \textit{red star reduction} rule.
\end{lemma}
\begin{proof}
Let $G$ be a connected \bumpy{} graph and let $\varrho = (H, R, \B)$ be any rule constructed for the pair $H, R$. Suppose that $(H, R, \B)$ applies to $G$. Let $\phi$ be the witnessing isomorphism.

Focus on the vertices in $L$. As $L \subseteq R$ we have that $\forall v \in L, N_G(\phi(v)) = \phi(C)$. But then sets $\phi(C)$ and $\phi(L)$ satisfy the conditions of the \textit{red star reduction} rule and therefore the \textit{red star reduction} rule applies to $G$.
\qed
\end{proof}

\subsection{Obtained Results}
With careful implementation the $(\F,\A,\beta)$-Algorithm together with our handmade reduction rules is able to achieve the results as summarized in \autoref{results-summary}.
Note that $\F$ is fixed to $\{P_d\}$, $\A$ is as described in the previous subsection and the only parameter that varies is $\beta$.
The question is then whether the algorithm finishes with the given $\beta$ or not.
If it does finish, then we obtained a correct algorithm of running time $\ostar(\gamma^k)$ for some $\gamma \le \beta$.
The table contains, for each $d$, the least values of $\beta$ (or rather $\gamma$) for which our implementation of the algorithm finished.
The full source code of the implementation is available at \url{https://github.com/generating-algorithms/generating-dpvc}.
%
As you can see, sadly, we were not able to improve the running time of \textsc{$2$-PVC}, but we do not know whether it is a limitation of the algorithm itself or a limitation of time, space, and resources.

To better understand the behavior of the generating algorithm, we provide plots of the number of branching rules and time it takes to achieve target branching factor.
The plots are obtained by subsequently running the algorithm with $\beta = \gamma - 0.005$ where $\ostar(\gamma^k)$ is the running time of the algorithm obtained in the previous run.
The runs depicted in the plots were performed on a virtual super-computer with 255 CPU cores and 128GB of RAM hosted within HPE Superdome Flex supercomputer (576 CPU threads provided by Intel Xeon Gold CPUs up to 4 GHz, 6 TB DDR4-2933 MHz RAM).
As the plots are not meant for a comparison with other algorithms, we only performed a single run for each branching factor.

Interestingly, sometimes better branching factors can be achieved faster. This can be caused, e.g., by obtaining less good rules in the early stages of the algorithm which speeds up the filtering of the subgraphs in the later stages.

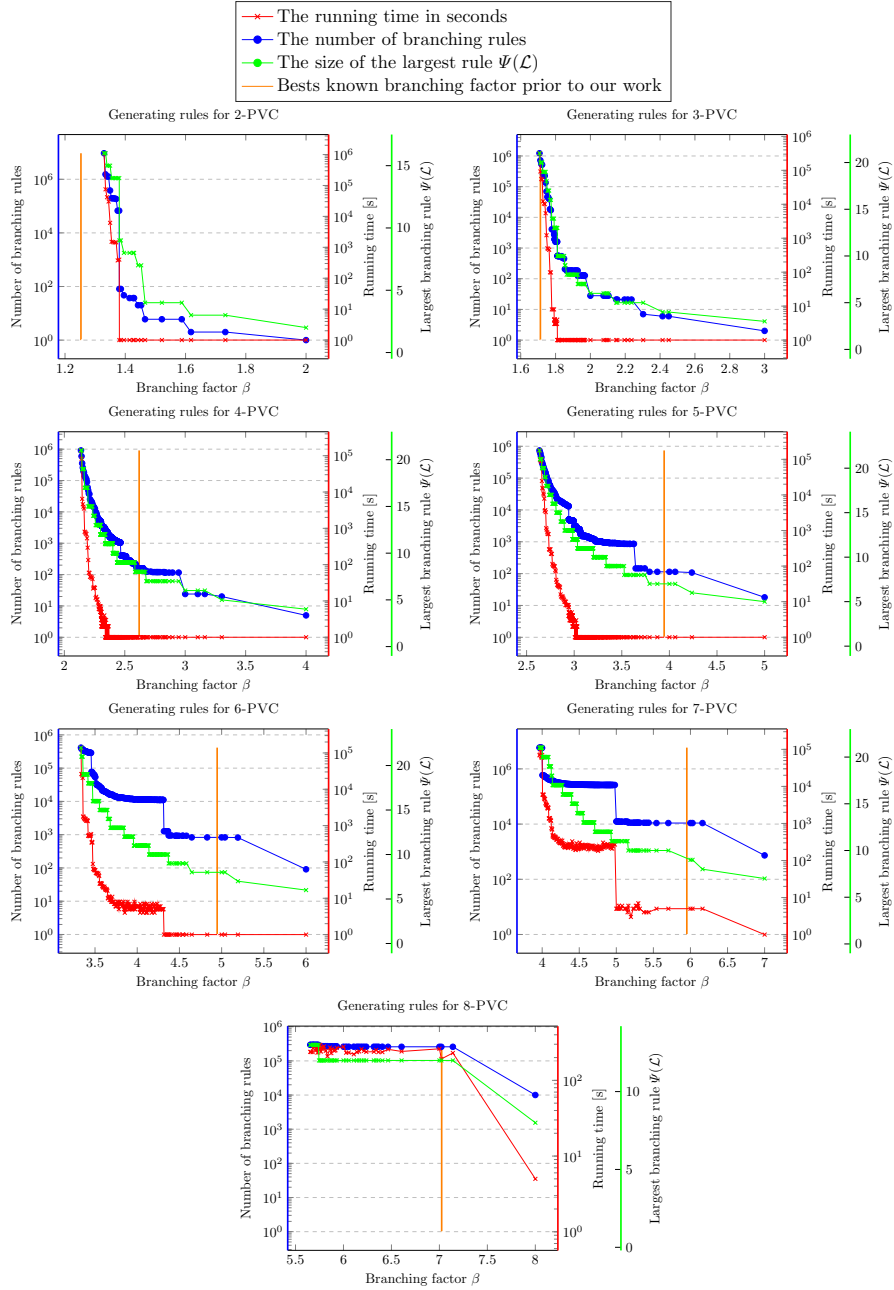
\begin{center}
\begin{minipage}[h]{0.99\textwidth}
\begin{center}
\resizebox {\textwidth/2} {!} {
\begin{tikzpicture}
    \begin{axis}[%
    hide axis,
    xmin=10,
    xmax=50,
    ymin=0,
    ymax=0.4,
    legend style={draw=white!15!black,legend cell align=left}
    ]
    \addlegendimage{red,mark=x}
    \addlegendentry{The running time in seconds};
    \addlegendimage{blue,mark=*}
    \addlegendentry{The number of branching rules};
    \addlegendimage{green,mark=*}
    \addlegendentry{The size of the largest rule $\Psi(\Llist)$};
    \addlegendimage{orange,mark=none}
    \addlegendentry{Bests known branching factor prior to our work};
    \end{axis}
\end{tikzpicture}
}
\end{center}
\end{minipage}
\begin{minipage}[t]{0.49\textwidth}
\resizebox {0.99\textwidth} {!} {
\begin{tikzpicture}
\begin{axis}[
    axis y line*=left,
    y axis line style={color=blue, very thick},
    title={Generating rules for \textsc{$2$-PVC$ $}},
    xlabel={Branching factor $\beta$},
    ylabel={Number of branching rules},
    ymode=log,
    legend pos=north west,
    ymajorgrids=true,
    grid style=dashed,
]

\addplot[color=blue,mark=*]
    coordinates {(1.32939,9345243)(1.3344,1527655)(1.33941,1270685)(1.3445,1254700)(1.34946,381811)(1.35446,195640)(1.3595,194366)(1.36461,192816)(1.36965,184258)(1.37472,67250)(1.3798,67250)(1.38028,81)(1.38534,81)(1.39534,47)(1.41422,37)(1.42623,37)(1.42912,37)(1.44226,20)(1.45263,20)(1.46558,6)(1.52139,6)(1.58741,6)(1.61804,2)(1.73206,2)(2,1)};
    \legend{}

\addplot [color=orange,mark=none, very thick] coordinates {(1.25288, 1) (1.25288, 9345243)};

\end{axis}

\begin{axis}[
  axis y line*=right,
  y axis line style={color=red, very thick},
  axis x line=none,
  ymode=log,
  ylabel={Running time [s]},
]
\addplot[color=red,mark=x]
    coordinates {(1.32939,1107514)(1.3344,75194)(1.33941,41670)(1.3445,30318)(1.34946,6194)(1.35446,1509)(1.3595,1470)(1.36461,1423)(1.36965,1446)(1.37472,377)(1.3798,400)(1.38028,1)(1.38534,1)(1.39534,1)(1.41422,1)(1.42623,1)(1.42912,1)(1.44226,1)(1.45263,1)(1.46558,1)(1.52139,1)(1.58741,1)(1.61804,1)(1.73206,1)(2,1)};

\addplot [color=white,mark=none] coordinates {(1.25288, 1) (1.25288, 1)};

\end{axis}

\begin{axis}[
  axis y line*=right,
  y axis line style={color=green, very thick},
  axis x line=none,
  ylabel={Largest branching rule $\Psi(\Llist)$},
]
\pgfplotsset{every outer y axis line/.style={xshift=1.6cm, color=green, very thick}, every tick/.style={xshift=1.6cm}, every y tick label/.style={xshift=1.6cm} }

\addplot[color=green,mark=x]
    coordinates {(1.32939,16)(1.3344,16)(1.33941,15)(1.3445,15)(1.34946,15)(1.35446,14)(1.3595,14)(1.36461,14)(1.36965,14)(1.37472,14)(1.3798,14)(1.38028,9)(1.38534,9)(1.39534,8)(1.41422,8)(1.42623,8)(1.42912,8)(1.44226,7)(1.45263,7)(1.46558,4)(1.52139,4)(1.58741,4)(1.61804,3)(1.73206,3)(2,2)};

\addplot [color=white,mark=none] coordinates {(1.25288, 1) (1.25288, 1)};

\end{axis}

\end{tikzpicture}
}
\end{minipage}
\begin{minipage}[t]{0.49\textwidth}
\resizebox {0.99\textwidth} {!} {
\begin{tikzpicture}
\begin{axis}[
    axis y line*=left,
    y axis line style={color=blue, very thick},
    title={Generating rules for \textsc{$3$-PVC$ $}},
    xlabel={Branching factor $\beta$},
    ylabel={Number of branching rules},
    ymode=log,
    legend pos=north west,
    ymajorgrids=true,
    grid style=dashed,
]

\addplot[color=blue,mark=*]
    coordinates {(1.70779,1226384)(1.71279,722098)(1.71784,548593)(1.7229,518081)(1.72785,259042)(1.7329,227150)(1.73789,226174)(1.74288,134814)(1.74789,69787)(1.75288,46379)(1.75787,45062)(1.76286,39423)(1.7679,18490)(1.77307,17155)(1.77829,4085)(1.78538,3958)(1.79107,3958)(1.79204,2768)(1.7934,2768)(1.79394,2768)(1.79411,2768)(1.79432,1934)(1.79519,1934)(1.7953,1934)(1.79633,1849)(1.79892,1604)(1.80241,1604)(1.8042,1604)(1.80844,1604)(1.80946,1604)(1.81054,552)(1.81166,552)(1.81241,552)(1.81298,552)(1.81713,552)(1.81822,552)(1.81894,544)(1.82175,544)(1.82398,544)(1.83509,544)(1.83929,478)(1.84891,452)(1.85357,206)(1.85615,206)(1.86371,184)(1.87372,184)(1.87573,184)(1.88186,184)(1.88852,184)(1.89081,184)(1.8933,184)(1.89933,184)(1.90898,184)(1.9133,184)(1.91719,184)(1.92371,184)(1.9267,184)(1.92757,127)(1.92945,127)(1.93199,127)(1.93709,127)(1.9498,127)(1.95459,127)(1.95656,127)(1.95844,127)(1.9601,127)(1.96168,127)(1.96315,127)(1.9645,127)(1.968,127)(2,28)(2.07549,28)(2.08675,28)(2.09456,28)(2.10381,28)(2.1479,21)(2.15444,21)(2.19583,21)(2.21433,21)(2.23608,21)(2.30278,7)(2.41422,6)(2.4495,6)(3,2)};
    \legend{}

\addplot [color=orange,mark=none, very thick] coordinates {(1.713, 1) (1.713, 1226384)};

\end{axis}

\begin{axis}[
  axis y line*=right,
  y axis line style={color=red, very thick},
  axis x line=none,
  ymode=log,
  ylabel={Running time [s]},
]
\addplot[color=red,mark=x]
    coordinates {(1.70779,319679)(1.71279,92652)(1.71784,55465)(1.7229,53340)(1.72785,12189)(1.7329,10499)(1.73789,10112)(1.74288,5460)(1.74789,1224)(1.75288,504)(1.75787,487)(1.76286,456)(1.7679,99)(1.77307,98)(1.77829,8)(1.78538,8)(1.79107,8)(1.79204,4)(1.7934,4)(1.79394,3)(1.79411,4)(1.79432,3)(1.79519,4)(1.7953,4)(1.79633,3)(1.79892,3)(1.80241,3)(1.8042,3)(1.80844,3)(1.80946,4)(1.81054,1)(1.81166,1)(1.81241,1)(1.81298,1)(1.81713,1)(1.81822,1)(1.81894,1)(1.82175,1)(1.82398,1)(1.83509,1)(1.83929,1)(1.84891,1)(1.85357,1)(1.85615,1)(1.86371,1)(1.87372,1)(1.87573,1)(1.88186,1)(1.88852,1)(1.89081,1)(1.8933,1)(1.89933,1)(1.90898,1)(1.9133,1)(1.91719,1)(1.92371,1)(1.9267,1)(1.92757,1)(1.92945,1)(1.93199,1)(1.93709,1)(1.9498,1)(1.95459,1)(1.95656,1)(1.95844,1)(1.9601,1)(1.96168,1)(1.96315,1)(1.9645,1)(1.968,1)(2,1)(2.07549,1)(2.08675,1)(2.09456,1)(2.10381,1)(2.1479,1)(2.15444,1)(2.19583,1)(2.21433,1)(2.23608,1)(2.30278,1)(2.41422,1)(2.4495,1)(3,1)};

\addplot [color=white,mark=none] coordinates {(1.713, 1) (1.713, 1)};

\end{axis}

\begin{axis}[
  axis y line*=right,
  y axis line style={color=green, very thick},
  axis x line=none,
  ylabel={Largest branching rule $\Psi(\Llist)$},
]
\pgfplotsset{every outer y axis line/.style={xshift=1.6cm, color=green, very thick}, every tick/.style={xshift=1.6cm}, every y tick label/.style={xshift=1.6cm} }

\addplot[color=green,mark=x]
    coordinates {(1.70779,21)(1.71279,20)(1.71784,20)(1.7229,20)(1.72785,19)(1.7329,19)(1.73789,19)(1.74288,19)(1.74789,18)(1.75288,17)(1.75787,17)(1.76286,17)(1.7679,16)(1.77307,16)(1.77829,14)(1.78538,14)(1.79107,14)(1.79204,13)(1.7934,13)(1.79394,13)(1.79411,13)(1.79432,13)(1.79519,13)(1.7953,13)(1.79633,13)(1.79892,13)(1.80241,13)(1.8042,13)(1.80844,13)(1.80946,13)(1.81054,10)(1.81166,10)(1.81241,10)(1.81298,10)(1.81713,10)(1.81822,10)(1.81894,10)(1.82175,10)(1.82398,10)(1.83509,10)(1.83929,10)(1.84891,10)(1.85357,9)(1.85615,9)(1.86371,8)(1.87372,8)(1.87573,8)(1.88186,8)(1.88852,8)(1.89081,8)(1.8933,8)(1.89933,8)(1.90898,8)(1.9133,8)(1.91719,8)(1.92371,8)(1.9267,8)(1.92757,7)(1.92945,7)(1.93199,7)(1.93709,7)(1.9498,7)(1.95459,7)(1.95656,7)(1.95844,7)(1.9601,7)(1.96168,7)(1.96315,7)(1.9645,7)(1.968,7)(2,6)(2.07549,6)(2.08675,6)(2.09456,6)(2.10381,6)(2.1479,5)(2.15444,5)(2.19583,5)(2.21433,5)(2.23608,5)(2.30278,5)(2.41422,4)(2.4495,4)(3,3)};

\addplot [color=white,mark=none] coordinates {(1.713, 1) (1.713, 1)};

\end{axis}

\end{tikzpicture}
}
\end{minipage}
\begin{minipage}[t]{0.49\textwidth}
\resizebox {0.99\textwidth} {!} {
\input{plots/4pvc-bf-brcnt-times-plot.tex}
}
\end{minipage}
\begin{minipage}[t]{0.49\textwidth}
\resizebox {0.99\textwidth} {!} {
\input{plots/5pvc-bf-brcnt-times-plot.tex}
}
\end{minipage}
\begin{minipage}[t]{0.49\textwidth}
\resizebox {0.99\textwidth} {!} {
\begin{tikzpicture}
\begin{axis}[
    axis y line*=left,
    y axis line style={color=blue, very thick},
    title={Generating rules for \textsc{$6$-PVC$ $}},
    xlabel={Branching factor $\beta$},
    ylabel={Number of branching rules},
    ymode=log,
    legend pos=north west,
    ymajorgrids=true,
    grid style=dashed,
]

\addplot[color=blue,mark=*]
    coordinates {(3.33339,414247)(3.33839,385994)(3.34339,378806)(3.34838,376404)(3.3534,366834)(3.35838,347557)(3.36335,343867)(3.36836,336660)(3.37337,333555)(3.3784,329007)(3.38337,324529)(3.38844,320967)(3.39349,316826)(3.39847,314663)(3.4035,312834)(3.40848,305765)(3.41347,303726)(3.4186,299539)(3.42355,296107)(3.42859,294504)(3.43359,292252)(3.4386,291554)(3.44358,290598)(3.44858,289375)(3.45357,288768)(3.45859,77584)(3.46359,75089)(3.46858,74743)(3.47363,67820)(3.47868,64900)(3.48369,63618)(3.48872,62823)(3.49376,62371)(3.49877,54989)(3.50375,53880)(3.50876,34801)(3.51389,34478)(3.51888,31699)(3.52389,31076)(3.52901,31012)(3.53405,30542)(3.53914,30478)(3.54428,29763)(3.54938,29011)(3.55441,28518)(3.55949,26468)(3.56456,26133)(3.56975,25994)(3.57479,25863)(3.57975,24689)(3.58478,21394)(3.58983,21345)(3.59527,21318)(3.60037,21045)(3.60536,20359)(3.61037,20130)(3.61547,19676)(3.62048,19645)(3.6255,19530)(3.63066,19530)(3.63575,18772)(3.641,18453)(3.64635,17991)(3.65144,17531)(3.65684,17531)(3.66195,17531)(3.66707,16426)(3.67215,16255)(3.67725,16255)(3.68252,16044)(3.68775,15943)(3.69329,15912)(3.6984,15896)(3.70422,15889)(3.70973,15851)(3.71494,15120)(3.72005,15120)(3.72546,14392)(3.73063,14370)(3.73743,14367)(3.74271,14367)(3.74802,13934)(3.75305,13934)(3.7589,13747)(3.76457,13268)(3.77073,13268)(3.77623,13268)(3.78184,13268)(3.78757,13263)(3.79377,13233)(3.80139,13233)(3.80663,12941)(3.81264,12941)(3.81838,12941)(3.82478,12941)(3.83006,12808)(3.83538,12808)(3.84039,12670)(3.84689,12595)(3.85216,12567)(3.85753,12454)(3.86327,12446)(3.86842,12446)(3.87392,12410)(3.87906,12322)(3.88508,12251)(3.89103,12251)(3.89628,12251)(3.90188,12124)(3.90886,11884)(3.91424,11884)(3.92069,11873)(3.92814,11873)(3.93396,11873)(3.94051,11873)(3.94603,11873)(3.95138,11635)(3.95935,11606)(3.96626,11606)(3.96841,11606)(3.96972,11606)(3.97115,11606)(3.97395,11606)(4,11407)(4.00858,11407)(4.0124,11407)(4.01548,11407)(4.02056,11407)(4.02458,11407)(4.02912,11407)(4.03191,11407)(4.03405,11407)(4.03656,11407)(4.03938,11407)(4.04097,11407)(4.04669,11407)(4.04892,11407)(4.05765,11407)(4.05933,11407)(4.0611,11403)(4.06421,11403)(4.06725,11403)(4.06953,11403)(4.0722,11403)(4.075,11403)(4.0778,11403)(4.08061,11344)(4.08564,11344)(4.08808,11344)(4.08907,11344)(4.09065,11344)(4.09583,11332)(4.10295,11332)(4.11626,11332)(4.11795,11332)(4.11959,11332)(4.12356,11332)(4.13137,11332)(4.13556,11332)(4.14093,11293)(4.14454,11293)(4.14795,11293)(4.15633,11293)(4.16229,11250)(4.17136,11250)(4.17283,11250)(4.17576,11250)(4.19259,11229)(4.19971,11200)(4.20308,11200)(4.20482,11200)(4.20867,11200)(4.22418,11198)(4.22637,11198)(4.22898,11198)(4.23607,11121)(4.24265,11121)(4.24843,11121)(4.26058,11121)(4.26297,11121)(4.26664,11121)(4.27253,11121)(4.27376,11121)(4.27493,11121)(4.28764,11118)(4.29715,11118)(4.29929,11118)(4.30556,11118)(4.31076,11118)(4.31663,1270)(4.33692,1267)(4.3511,1267)(4.35891,1267)(4.36489,1267)(4.37229,949)(4.38418,949)(4.40513,949)(4.4495,929)(4.46411,929)(4.47214,929)(4.4945,929)(4.53787,929)(4.54139,929)(4.58258,929)(4.64576,829)(4.82844,829)(5,829)(5.03939,829)(5.19259,820)(6,91)};
    \legend{}

\addplot [color=orange,mark=none, very thick] coordinates {(4.947, 1) (4.947, 414247)};

\end{axis}

\begin{axis}[
  axis y line*=right,
  y axis line style={color=red, very thick},
  axis x line=none,
  ymode=log,
  ylabel={Running time [s]},
]
\addplot[color=red,mark=x]
    coordinates {(3.33339,139047)(3.33839,26208)(3.34339,26292)(3.34838,23976)(3.3534,20570)(3.35838,1781)(3.36335,1744)(3.36836,1600)(3.37337,1592)(3.3784,1505)(3.38337,1469)(3.38844,1533)(3.39349,1388)(3.39847,1360)(3.4035,1384)(3.40848,1391)(3.41347,1328)(3.4186,525)(3.42355,527)(3.42859,550)(3.43359,514)(3.4386,513)(3.44358,526)(3.44858,538)(3.45357,583)(3.45859,342)(3.46359,357)(3.46858,343)(3.47363,88)(3.47868,64)(3.48369,64)(3.48872,61)(3.49376,59)(3.49877,60)(3.50375,59)(3.50876,50)(3.51389,49)(3.51888,50)(3.52389,49)(3.52901,47)(3.53405,46)(3.53914,48)(3.54428,43)(3.54938,41)(3.55441,25)(3.55949,26)(3.56456,26)(3.56975,25)(3.57479,26)(3.57975,25)(3.58478,22)(3.58983,21)(3.59527,20)(3.60037,20)(3.60536,20)(3.61037,21)(3.61547,19)(3.62048,18)(3.6255,17)(3.63066,17)(3.63575,18)(3.641,17)(3.64635,16)(3.65144,10)(3.65684,10)(3.66195,11)(3.66707,11)(3.67215,9)(3.67725,11)(3.68252,9)(3.68775,9)(3.69329,8)(3.6984,8)(3.70422,11)(3.70973,8)(3.71494,8)(3.72005,8)(3.72546,7)(3.73063,7)(3.73743,7)(3.74271,6)(3.74802,6)(3.75305,6)(3.7589,7)(3.76457,8)(3.77073,8)(3.77623,6)(3.78184,5)(3.78757,8)(3.79377,8)(3.80139,7)(3.80663,7)(3.81264,7)(3.81838,6)(3.82478,7)(3.83006,7)(3.83538,8)(3.84039,5)(3.84689,6)(3.85216,4)(3.85753,5)(3.86327,5)(3.86842,6)(3.87392,7)(3.87906,6)(3.88508,8)(3.89103,7)(3.89628,5)(3.90188,5)(3.90886,7)(3.91424,5)(3.92069,5)(3.92814,6)(3.93396,5)(3.94051,5)(3.94603,6)(3.95138,5)(3.95935,6)(3.96626,8)(3.96841,7)(3.96972,5)(3.97115,6)(3.97395,5)(4,6)(4.00858,6)(4.0124,6)(4.01548,6)(4.02056,6)(4.02458,6)(4.02912,5)(4.03191,5)(4.03405,4)(4.03656,7)(4.03938,5)(4.04097,5)(4.04669,6)(4.04892,6)(4.05765,5)(4.05933,5)(4.0611,5)(4.06421,5)(4.06725,5)(4.06953,4)(4.0722,5)(4.075,7)(4.0778,5)(4.08061,6)(4.08564,6)(4.08808,7)(4.08907,6)(4.09065,6)(4.09583,7)(4.10295,4)(4.11626,5)(4.11795,4)(4.11959,7)(4.12356,6)(4.13137,5)(4.13556,5)(4.14093,6)(4.14454,5)(4.14795,6)(4.15633,5)(4.16229,5)(4.17136,7)(4.17283,4)(4.17576,5)(4.19259,6)(4.19971,6)(4.20308,7)(4.20482,4)(4.20867,5)(4.22418,5)(4.22637,6)(4.22898,5)(4.23607,7)(4.24265,5)(4.24843,5)(4.26058,6)(4.26297,5)(4.26664,5)(4.27253,5)(4.27376,5)(4.27493,6)(4.28764,5)(4.29715,5)(4.29929,5)(4.30556,5)(4.31076,5)(4.31663,1)(4.33692,1)(4.3511,1)(4.35891,1)(4.36489,1)(4.37229,1)(4.38418,1)(4.40513,1)(4.4495,1)(4.46411,1)(4.47214,1)(4.4945,1)(4.53787,1)(4.54139,1)(4.58258,1)(4.64576,1)(4.82844,1)(5,1)(5.03939,1)(5.19259,1)(6,1)};

\addplot [color=white,mark=none] coordinates {(4.947, 1) (4.947, 1)};

\end{axis}

\begin{axis}[
  axis y line*=right,
  y axis line style={color=green, very thick},
  axis x line=none,
  ylabel={Largest branching rule $\Psi(\Llist)$},
]
\pgfplotsset{every outer y axis line/.style={xshift=1.6cm, color=green, very thick}, every tick/.style={xshift=1.6cm}, every y tick label/.style={xshift=1.6cm} }

\addplot[color=green,mark=x]
    coordinates {(3.33339,22)(3.33839,21)(3.34339,21)(3.34838,21)(3.3534,21)(3.35838,19)(3.36335,19)(3.36836,19)(3.37337,19)(3.3784,19)(3.38337,19)(3.38844,19)(3.39349,19)(3.39847,19)(3.4035,19)(3.40848,19)(3.41347,19)(3.4186,18)(3.42355,18)(3.42859,18)(3.43359,18)(3.4386,18)(3.44358,18)(3.44858,18)(3.45357,18)(3.45859,18)(3.46359,18)(3.46858,18)(3.47363,16)(3.47868,16)(3.48369,16)(3.48872,16)(3.49376,16)(3.49877,16)(3.50375,16)(3.50876,16)(3.51389,16)(3.51888,16)(3.52389,16)(3.52901,16)(3.53405,16)(3.53914,16)(3.54428,16)(3.54938,16)(3.55441,15)(3.55949,15)(3.56456,15)(3.56975,15)(3.57479,15)(3.57975,15)(3.58478,15)(3.58983,15)(3.59527,15)(3.60037,15)(3.60536,15)(3.61037,15)(3.61547,15)(3.62048,15)(3.6255,15)(3.63066,15)(3.63575,15)(3.641,15)(3.64635,15)(3.65144,14)(3.65684,14)(3.66195,14)(3.66707,14)(3.67215,14)(3.67725,14)(3.68252,13)(3.68775,13)(3.69329,13)(3.6984,13)(3.70422,13)(3.70973,13)(3.71494,13)(3.72005,13)(3.72546,13)(3.73063,13)(3.73743,13)(3.74271,13)(3.74802,13)(3.75305,13)(3.7589,13)(3.76457,13)(3.77073,13)(3.77623,13)(3.78184,13)(3.78757,13)(3.79377,13)(3.80139,13)(3.80663,13)(3.81264,13)(3.81838,13)(3.82478,13)(3.83006,13)(3.83538,13)(3.84039,13)(3.84689,12)(3.85216,12)(3.85753,12)(3.86327,12)(3.86842,12)(3.87392,12)(3.87906,12)(3.88508,12)(3.89103,12)(3.89628,12)(3.90188,12)(3.90886,12)(3.91424,12)(3.92069,12)(3.92814,12)(3.93396,12)(3.94051,12)(3.94603,12)(3.95138,12)(3.95935,11)(3.96626,11)(3.96841,11)(3.96972,11)(3.97115,11)(3.97395,11)(4,11)(4.00858,11)(4.0124,11)(4.01548,11)(4.02056,11)(4.02458,11)(4.02912,11)(4.03191,11)(4.03405,11)(4.03656,11)(4.03938,11)(4.04097,11)(4.04669,11)(4.04892,11)(4.05765,11)(4.05933,11)(4.0611,11)(4.06421,11)(4.06725,11)(4.06953,11)(4.0722,11)(4.075,11)(4.0778,11)(4.08061,11)(4.08564,11)(4.08808,11)(4.08907,11)(4.09065,11)(4.09583,11)(4.10295,11)(4.11626,11)(4.11795,11)(4.11959,11)(4.12356,11)(4.13137,11)(4.13556,11)(4.14093,10)(4.14454,10)(4.14795,10)(4.15633,10)(4.16229,10)(4.17136,10)(4.17283,10)(4.17576,10)(4.19259,10)(4.19971,10)(4.20308,10)(4.20482,10)(4.20867,10)(4.22418,10)(4.22637,10)(4.22898,10)(4.23607,10)(4.24265,10)(4.24843,10)(4.26058,10)(4.26297,10)(4.26664,10)(4.27253,10)(4.27376,10)(4.27493,10)(4.28764,10)(4.29715,10)(4.29929,10)(4.30556,10)(4.31076,10)(4.31663,10)(4.33692,10)(4.3511,10)(4.35891,10)(4.36489,10)(4.37229,9)(4.38418,9)(4.40513,9)(4.4495,9)(4.46411,9)(4.47214,9)(4.4945,9)(4.53787,9)(4.54139,9)(4.58258,9)(4.64576,8)(4.82844,8)(5,8)(5.03939,8)(5.19259,7)(6,6)};

\addplot [color=white,mark=none] coordinates {(4.947, 1) (4.947, 1)};

\end{axis}

\end{tikzpicture}
}
\end{minipage}
\begin{minipage}[t]{0.49\textwidth}
\resizebox {0.99\textwidth} {!} {
\begin{tikzpicture}
\begin{axis}[
    axis y line*=left,
    y axis line style={color=blue, very thick},
    title={Generating rules for \textsc{$7$-PVC$ $}},
    xlabel={Branching factor $\beta$},
    ylabel={Number of branching rules},
    ymode=log,
    legend pos=north west,
    ymajorgrids=true,
    grid style=dashed,
]

\addplot[color=blue,mark=*]
    coordinates {(3.96376,5895811)(3.96877,5833018)(3.9738,5817171)(3.97879,5805354)(3.98379,5772449)(3.98877,5767908)(3.99379,5756372)(3.99876,5755765)(4.00399,592962)(4.009,587887)(4.01399,582068)(4.01897,576728)(4.02396,566768)(4.02898,555641)(4.03395,548691)(4.03899,514205)(4.04399,504691)(4.04903,498507)(4.05416,493535)(4.05919,489169)(4.06419,484224)(4.06919,479405)(4.07418,439323)(4.07924,427677)(4.08427,423120)(4.08929,419832)(4.09426,418700)(4.09929,416630)(4.1043,413519)(4.10926,406930)(4.1143,394030)(4.11929,388561)(4.12432,386137)(4.1294,384641)(4.1344,382782)(4.13937,378530)(4.14438,373377)(4.14944,366448)(4.15448,355238)(4.15953,349154)(4.1646,342938)(4.16964,341947)(4.17469,338844)(4.1797,334016)(4.18482,332838)(4.18992,329113)(4.19499,325206)(4.20004,324117)(4.2051,323355)(4.21018,319244)(4.21515,317230)(4.22019,313998)(4.22519,312029)(4.23031,308709)(4.23542,308705)(4.24067,307685)(4.24567,307436)(4.25067,306923)(4.25565,305622)(4.26069,299259)(4.26569,297914)(4.27075,295875)(4.27598,295092)(4.28105,294173)(4.28607,294087)(4.2911,293551)(4.29614,290831)(4.30118,290804)(4.30616,288420)(4.31133,286813)(4.3167,285563)(4.32169,284701)(4.32698,284686)(4.33203,283732)(4.33712,283221)(4.34239,282980)(4.34738,281456)(4.35244,281384)(4.35751,281375)(4.36263,280405)(4.36779,280382)(4.37287,279286)(4.3781,279286)(4.38339,278167)(4.38835,277882)(4.39374,277840)(4.39936,276479)(4.40461,276369)(4.41001,276182)(4.41524,276155)(4.4209,274484)(4.42596,274474)(4.43139,274444)(4.43654,271983)(4.44163,271983)(4.44674,271983)(4.45183,270641)(4.45727,270641)(4.46243,270633)(4.46801,270564)(4.47347,269226)(4.47878,269176)(4.48391,269030)(4.48977,268785)(4.49494,268763)(4.50037,268737)(4.506,268730)(4.51141,268517)(4.51691,268514)(4.52234,267757)(4.52761,267714)(4.53285,267714)(4.53859,267191)(4.54358,266913)(4.54937,266913)(4.55529,266842)(4.56079,265445)(4.56588,265445)(4.57202,265445)(4.57752,265269)(4.58284,265259)(4.58869,265203)(4.59632,265203)(4.60145,265203)(4.60657,265196)(4.61187,264151)(4.61927,264151)(4.62469,264108)(4.63165,264108)(4.63676,264108)(4.64224,264108)(4.64803,263443)(4.65447,263443)(4.65953,263435)(4.66492,263435)(4.67161,263422)(4.67684,263422)(4.68213,262898)(4.68891,262898)(4.69415,262898)(4.69916,262439)(4.70419,262153)(4.71063,262011)(4.71603,262011)(4.7212,262011)(4.729,261958)(4.7355,261902)(4.74121,261902)(4.7464,261889)(4.75348,261889)(4.7585,261889)(4.76645,261530)(4.77201,261530)(4.77882,261530)(4.78748,261530)(4.80018,261510)(4.80585,261510)(4.82126,261510)(4.82844,261364)(4.83874,261364)(4.84389,261364)(4.85411,261345)(4.86041,261345)(4.86714,261341)(4.87299,261341)(4.88054,261341)(4.88749,261334)(4.8925,261334)(4.90105,261313)(4.90675,261313)(4.91424,261313)(4.91969,261313)(4.92507,261313)(4.93395,261206)(4.94339,261206)(4.95045,261206)(4.96033,261206)(4.98028,261206)(5,12411)(5.00797,12411)(5.02481,12411)(5.03294,12257)(5.03938,12257)(5.06511,12257)(5.07758,12257)(5.11469,12257)(5.12312,12257)(5.14007,12257)(5.15078,12257)(5.16228,11279)(5.19616,11258)(5.22788,11258)(5.24265,11258)(5.26227,11258)(5.27492,11258)(5.29151,11258)(5.2958,11258)(5.31663,11258)(5.37229,11183)(5.40432,11183)(5.43564,11183)(5.54139,10907)(5.70157,10907)(6,10907)(6.02753,10907)(6.16229,10894)(7,734)};
    \legend{}

\addplot [color=orange,mark=none, very thick] coordinates {(5.951, 1) (5.951, 5895811)};

\end{axis}

\begin{axis}[
  axis y line*=right,
  y axis line style={color=red, very thick},
  axis x line=none,
  ymode=log,
  ylabel={Running time [s]},
]
\addplot[color=red,mark=x]
    coordinates {(3.96376,69288)(3.96877,72389)(3.9738,110756)(3.97879,99238)(3.98379,66030)(3.98877,61679)(3.99379,59455)(3.99876,70541)(4.00399,6050)(4.009,5915)(4.01399,5943)(4.01897,5973)(4.02396,4997)(4.02898,4558)(4.03395,4649)(4.03899,4552)(4.04399,3564)(4.04903,3364)(4.05416,3189)(4.05919,3159)(4.06419,3040)(4.06919,2662)(4.07418,2522)(4.07924,2753)(4.08427,2620)(4.08929,1365)(4.09426,1379)(4.09929,1339)(4.1043,1400)(4.10926,1259)(4.1143,1195)(4.11929,1189)(4.12432,750)(4.1294,697)(4.1344,690)(4.13937,475)(4.14438,438)(4.14944,483)(4.15448,458)(4.15953,419)(4.1646,431)(4.16964,401)(4.17469,444)(4.1797,412)(4.18482,416)(4.18992,367)(4.19499,387)(4.20004,365)(4.2051,386)(4.21018,382)(4.21515,327)(4.22019,342)(4.22519,330)(4.23031,311)(4.23542,300)(4.24067,293)(4.24567,347)(4.25067,305)(4.25565,300)(4.26069,300)(4.26569,285)(4.27075,344)(4.27598,290)(4.28105,252)(4.28607,253)(4.2911,278)(4.29614,238)(4.30118,232)(4.30616,224)(4.31133,255)(4.3167,254)(4.32169,229)(4.32698,240)(4.33203,251)(4.33712,230)(4.34239,274)(4.34738,217)(4.35244,230)(4.35751,260)(4.36263,236)(4.36779,258)(4.37287,277)(4.3781,233)(4.38339,251)(4.38835,259)(4.39374,264)(4.39936,263)(4.40461,329)(4.41001,233)(4.41524,251)(4.4209,207)(4.42596,218)(4.43139,234)(4.43654,236)(4.44163,220)(4.44674,198)(4.45183,192)(4.45727,223)(4.46243,292)(4.46801,212)(4.47347,245)(4.47878,220)(4.48391,206)(4.48977,217)(4.49494,232)(4.50037,204)(4.506,305)(4.51141,241)(4.51691,261)(4.52234,213)(4.52761,225)(4.53285,232)(4.53859,265)(4.54358,266)(4.54937,250)(4.55529,222)(4.56079,232)(4.56588,211)(4.57202,261)(4.57752,249)(4.58284,238)(4.58869,249)(4.59632,257)(4.60145,251)(4.60657,227)(4.61187,261)(4.61927,201)(4.62469,251)(4.63165,210)(4.63676,231)(4.64224,239)(4.64803,230)(4.65447,214)(4.65953,250)(4.66492,238)(4.67161,250)(4.67684,232)(4.68213,251)(4.68891,231)(4.69415,222)(4.69916,209)(4.70419,204)(4.71063,220)(4.71603,211)(4.7212,198)(4.729,227)(4.7355,197)(4.74121,243)(4.7464,197)(4.75348,182)(4.7585,218)(4.76645,235)(4.77201,202)(4.77882,237)(4.78748,204)(4.80018,235)(4.80585,207)(4.82126,304)(4.82844,262)(4.83874,221)(4.84389,263)(4.85411,235)(4.86041,245)(4.86714,259)(4.87299,261)(4.88054,276)(4.88749,275)(4.8925,231)(4.90105,260)(4.90675,261)(4.91424,257)(4.91969,229)(4.92507,217)(4.93395,264)(4.94339,209)(4.95045,243)(4.96033,279)(4.98028,270)(5,5)(5.00797,5)(5.02481,5)(5.03294,6)(5.03938,5)(5.06511,5)(5.07758,6)(5.11469,5)(5.12312,5)(5.14007,5)(5.15078,4)(5.16228,6)(5.19616,3)(5.22788,5)(5.24265,5)(5.26227,6)(5.27492,6)(5.29151,5)(5.2958,7)(5.31663,5)(5.37229,4)(5.40432,4)(5.43564,4)(5.54139,5)(5.70157,5)(6,5)(6.02753,5)(6.16229,5)(7,1)};

\addplot [color=white,mark=none] coordinates {(5.951, 1) (5.951, 1)};

\end{axis}

\begin{axis}[
  axis y line*=right,
  y axis line style={color=green, very thick},
  axis x line=none,
  ylabel={Largest branching rule $\Psi(\Llist)$},
]
\pgfplotsset{every outer y axis line/.style={xshift=1.6cm, color=green, very thick}, every tick/.style={xshift=1.6cm}, every y tick label/.style={xshift=1.6cm} }

\addplot[color=green,mark=x]
    coordinates {(3.96376,21)(3.96877,21)(3.9738,21)(3.97879,21)(3.98379,21)(3.98877,21)(3.99379,21)(3.99876,21)(4.00399,20)(4.009,20)(4.01399,20)(4.01897,20)(4.02396,20)(4.02898,20)(4.03395,20)(4.03899,20)(4.04399,20)(4.04903,20)(4.05416,20)(4.05919,20)(4.06419,20)(4.06919,20)(4.07418,20)(4.07924,20)(4.08427,20)(4.08929,19)(4.09426,19)(4.09929,19)(4.1043,19)(4.10926,19)(4.1143,19)(4.11929,19)(4.12432,18)(4.1294,18)(4.1344,18)(4.13937,17)(4.14438,17)(4.14944,17)(4.15448,17)(4.15953,17)(4.1646,17)(4.16964,17)(4.17469,17)(4.1797,17)(4.18482,17)(4.18992,17)(4.19499,17)(4.20004,17)(4.2051,17)(4.21018,17)(4.21515,17)(4.22019,17)(4.22519,17)(4.23031,17)(4.23542,17)(4.24067,17)(4.24567,17)(4.25067,17)(4.25565,17)(4.26069,17)(4.26569,17)(4.27075,17)(4.27598,16)(4.28105,16)(4.28607,16)(4.2911,16)(4.29614,16)(4.30118,16)(4.30616,16)(4.31133,16)(4.3167,16)(4.32169,16)(4.32698,16)(4.33203,16)(4.33712,16)(4.34239,16)(4.34738,16)(4.35244,16)(4.35751,16)(4.36263,16)(4.36779,16)(4.37287,16)(4.3781,16)(4.38339,16)(4.38835,16)(4.39374,16)(4.39936,15)(4.40461,15)(4.41001,15)(4.41524,15)(4.4209,15)(4.42596,15)(4.43139,15)(4.43654,15)(4.44163,15)(4.44674,15)(4.45183,15)(4.45727,15)(4.46243,15)(4.46801,15)(4.47347,14)(4.47878,14)(4.48391,14)(4.48977,14)(4.49494,14)(4.50037,14)(4.506,14)(4.51141,14)(4.51691,14)(4.52234,14)(4.52761,14)(4.53285,14)(4.53859,14)(4.54358,14)(4.54937,14)(4.55529,13)(4.56079,13)(4.56588,13)(4.57202,13)(4.57752,13)(4.58284,13)(4.58869,13)(4.59632,13)(4.60145,13)(4.60657,13)(4.61187,13)(4.61927,13)(4.62469,13)(4.63165,13)(4.63676,13)(4.64224,13)(4.64803,13)(4.65447,13)(4.65953,13)(4.66492,13)(4.67161,13)(4.67684,13)(4.68213,13)(4.68891,13)(4.69415,13)(4.69916,13)(4.70419,13)(4.71063,12)(4.71603,12)(4.7212,12)(4.729,12)(4.7355,12)(4.74121,12)(4.7464,12)(4.75348,12)(4.7585,12)(4.76645,12)(4.77201,12)(4.77882,12)(4.78748,12)(4.80018,12)(4.80585,12)(4.82126,12)(4.82844,12)(4.83874,12)(4.84389,12)(4.85411,12)(4.86041,12)(4.86714,12)(4.87299,12)(4.88054,12)(4.88749,12)(4.8925,12)(4.90105,12)(4.90675,12)(4.91424,12)(4.91969,12)(4.92507,12)(4.93395,11)(4.94339,11)(4.95045,11)(4.96033,11)(4.98028,11)(5,11)(5.00797,11)(5.02481,11)(5.03294,11)(5.03938,11)(5.06511,11)(5.07758,11)(5.11469,11)(5.12312,11)(5.14007,11)(5.15078,11)(5.16228,10)(5.19616,10)(5.22788,10)(5.24265,10)(5.26227,10)(5.27492,10)(5.29151,10)(5.2958,10)(5.31663,10)(5.37229,10)(5.40432,10)(5.43564,10)(5.54139,10)(5.70157,10)(6,9)(6.02753,9)(6.16229,8)(7,7)};

\addplot [color=white,mark=none] coordinates {(5.951, 1) (5.951, 1)};

\end{axis}

\end{tikzpicture}
}
\end{minipage}
\begin{minipage}[t]{0.49\textwidth}
\resizebox {0.99\textwidth} {!} {
\begin{tikzpicture}
\begin{axis}[
    axis y line*=left,
    y axis line style={color=blue, very thick},
    title={Generating rules for \textsc{$8$-PVC$ $}},
    xlabel={Branching factor $\beta$},
    ylabel={Number of branching rules},
    ymode=log,
    legend pos=north west,
    ymajorgrids=true,
    grid style=dashed,
]

\addplot[color=blue,mark=*]
    coordinates {(5.65332,296044)(5.65907,296025)(5.67742,296025)(5.68275,296025)(5.69043,296025)(5.70157,295332)(5.71113,295332)(5.72016,295332)(5.72872,295332)(5.73806,295332)(5.74457,269888)(5.75539,269888)(5.76478,269888)(5.77201,269888)(5.78161,269888)(5.79584,269888)(5.81508,269888)(5.83276,269888)(5.85845,267296)(5.87299,267296)(5.87937,267042)(5.90423,267042)(5.91609,267042)(5.93033,267042)(6,260279)(6.02753,260279)(6.05457,260279)(6.10725,260279)(6.14007,260279)(6.16229,260140)(6.18773,260140)(6.21278,260140)(6.23744,260140)(6.31663,260088)(6.34033,260088)(6.36368,260088)(6.40513,260088)(6.46411,258889)(6.60556,258889)(7,258797)(7.0203,258797)(7.14007,258780)(8,10030)};
    \legend{}

\addplot [color=orange,mark=none, very thick] coordinates {(7.0237, 1) (7.0237, 296044)};

\end{axis}

\begin{axis}[
  axis y line*=right,
  y axis line style={color=red, very thick},
  axis x line=none,
  ymode=log,
  ylabel={Running time [s]},
]
\addplot[color=red,mark=x]
    coordinates {(5.65332,238)(5.65907,237)(5.67742,236)(5.68275,266)(5.69043,284)(5.70157,284)(5.71113,248)(5.72016,260)(5.72872,232)(5.73806,279)(5.74457,256)(5.75539,294)(5.76478,240)(5.77201,275)(5.78161,259)(5.79584,281)(5.81508,246)(5.83276,208)(5.85845,271)(5.87299,229)(5.87937,261)(5.90423,246)(5.91609,250)(5.93033,269)(6,278)(6.02753,232)(6.05457,233)(6.10725,221)(6.14007,236)(6.16229,241)(6.18773,259)(6.21278,245)(6.23744,238)(6.31663,239)(6.34033,251)(6.36368,236)(6.40513,236)(6.46411,257)(6.60556,241)(7,261)(7.0203,192)(7.14007,230)(8,5)};

\addplot [color=white,mark=none] coordinates {(7.0237, 1) (7.0237, 1)};

\end{axis}

\begin{axis}[
  axis y line*=right,
  y axis line style={color=green, very thick},
  axis x line=none,
  ylabel={Largest branching rule $\Psi(\Llist)$},
]
\pgfplotsset{every outer y axis line/.style={xshift=1.6cm, color=green, very thick}, every tick/.style={xshift=1.6cm}, every y tick label/.style={xshift=1.6cm} }

\addplot[color=green,mark=x]
    coordinates {(5.65332,13)(5.65907,13)(5.67742,13)(5.68275,13)(5.69043,13)(5.70157,13)(5.71113,13)(5.72016,13)(5.72872,13)(5.73806,13)(5.74457,12)(5.75539,12)(5.76478,12)(5.77201,12)(5.78161,12)(5.79584,12)(5.81508,12)(5.83276,12)(5.85845,12)(5.87299,12)(5.87937,12)(5.90423,12)(5.91609,12)(5.93033,12)(6,12)(6.02753,12)(6.05457,12)(6.10725,12)(6.14007,12)(6.16229,12)(6.18773,12)(6.21278,12)(6.23744,12)(6.31663,12)(6.34033,12)(6.36368,12)(6.40513,12)(6.46411,12)(6.60556,12)(7,12)(7.0203,12)(7.14007,12)(8,8)};

\addplot [color=white,mark=none] coordinates {(7.0237, 1) (7.0237, 1)};

\end{axis}

\end{tikzpicture}
}
\end{minipage}
\end{center}

\subsection{Influence of the Individual Parts of the Generating Algorithm on its Performance}

Since the framework consist of several functions with some of them independent of each other, it is interesting to examine how important are the individual parts in obtaining the achieved results.

We focus on the \textit{DominanceFree} and \textit{Adjust} functions and the handmade rules. By turning off some of them, we try to measure their importance.
The closest variant of our algorithm to the approach of Gramm et al.~\cite{GrammGHN04} is when we turn off the \textit{DominanceFree} and \textit{Adjust} functions, i.e., we use only handmade rules (though Gramm et al. used the hand made rules in somewhat different manner).

\begin{table}[h]
\centering
\caption{
    Comparison of the best runs of different variants of the algorithm with time limit of 90 minutes.
    The table shows best branching factor (bf) achieved,
    the number of rules needed (\#rules),
    and the time in seconds it took to achieve it.
    FA -- Full algorithm;
    HR, DF only -- Handmade rules with DominanceFree only;
    HR only -- Handmade rules only;
    FA, no HR -- Full algorithm without handmade rules;
    NONE -- everything turned off;
    }
\label{modifiers-comparsion-table}
\small
\setlength\tabcolsep{1mm}
\begin{tabular}{|l|l|r|r|l|r|r|l|r|r|}\hline
 & \multicolumn{3}{|c|}{ 3-\textsc{PVC} } & \multicolumn{3}{|c|}{ 4-\textsc{PVC} } & \multicolumn{3}{|c|}{ 5-\textsc{PVC} } \\\hline
			& bf & time[s] & \#rules & bf & time[s] & \#rules & bf & time[s] & \#rules \\\hline

FA 			& 1.799 &  412 & 1,849 	& 2.293 &  799 & 5,178 		& 2.85 &  2,552 & 19,743 \\\hline
HR, DF only \qquad	& 1.95 &  111 & 1,026 	& 2.37 &  2,375 & 12,008 	& 2.958 &  3,693 & 26,677 \\\hline
HR only 		& 2.303 &  $<1$ & 7 	& 3.0 &  $<1$ & 30 		& 3.646 &  2 & 373 \\\hline
FA, no HR 		& 1.799 &  398 & 1,849 	& 3.0 &  $<1$ & 24 		& 4.0 &  $<1$ & 114 \\\hline
NONE 			& 2.303 &  $<1$ & 7 	& 3.036 &  $<1$ & 31 		& 5.0 &  $<1$ & 18 \\\hline
\end{tabular}
\end{table}

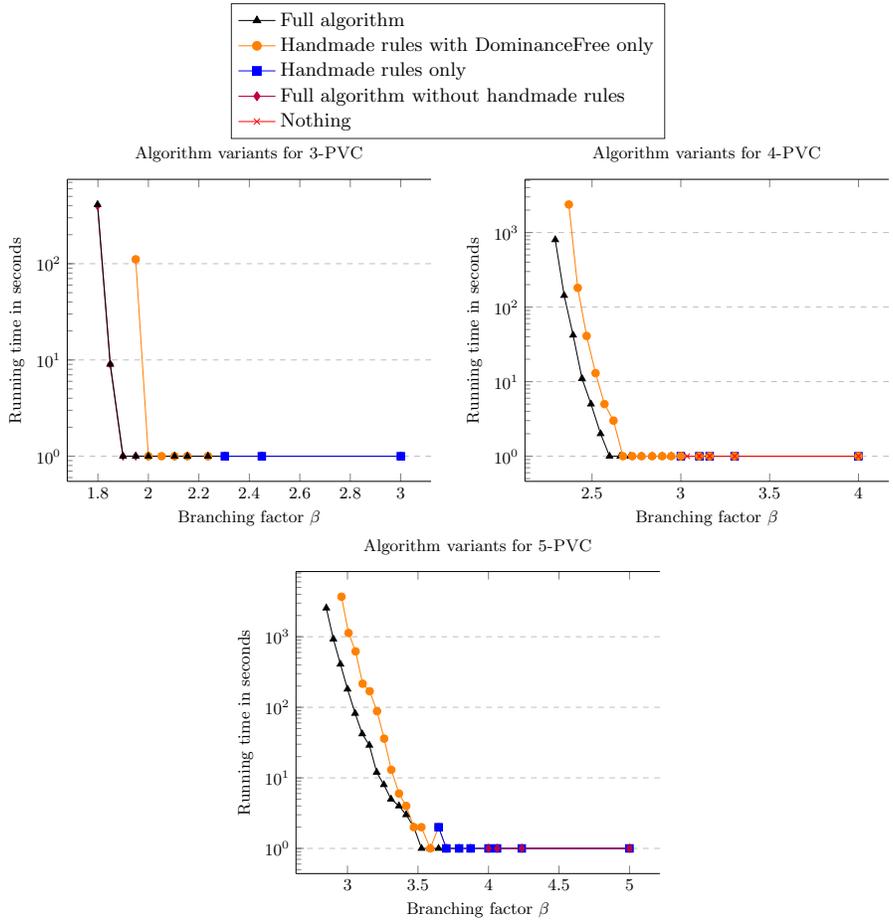
\begin{figure}[h!]

\begin{center}
\begin{minipage}[t]{0.99\textwidth}
\begin{center}
\resizebox {0.5\textwidth} {!} {
\begin{tikzpicture}
    \begin{axis}[%
    hide axis,
    xmin=10,
    xmax=50,
    ymin=0,
    ymax=0.4,
    legend style={draw=white!15!black,legend cell align=left}
    ]
    \addlegendimage{ black,mark=triangle* }
    \addlegendentry{ Full algorithm};

    \addlegendimage{ orange,mark=otimes* }
    \addlegendentry{ Handmade rules with DominanceFree only};

    \addlegendimage{ blue,mark=square* }
    \addlegendentry{ Handmade rules only};

    \addlegendimage{ purple,mark=diamond* }
    \addlegendentry{ Full algorithm without handmade rules};

    \addlegendimage{ red,mark=x }
    \addlegendentry{ Nothing};

    \end{axis}
\end{tikzpicture}
}
\end{center}
\end{minipage}
\begin{minipage}[t]{0.49\textwidth}
\resizebox {0.99\textwidth} {!} {
\begin{tikzpicture}
\begin{axis}[
    axis y line*=left,
    title={Algorithm variants for \textsc{$3$-PVC}},
    xlabel={Branching factor $\beta$},
    ylabel={Running time in seconds},
    ymode=log,
    legend pos=north west,
    ymajorgrids=true,
    grid style=dashed,
]

\addplot[color=purple,mark=diamond*]
    coordinates {(1.79868,398)(1.84891,9)(1.89933,1)(1.9498,1)(2,1)(2.10381,1)(2.15444,1)(2.23608,1)(2.30278,1)(2.4495,1)(3,1)};
    \legend{}

\addplot[color=orange,mark=otimes*]
    coordinates {(1.9498,111)(2,1)(2.05231,1)(2.10381,1)(2.15444,1)(2.23608,1)(2.30278,1)(2.4495,1)(3,1)};
    \legend{}

\addplot[color=red,mark=x]
    coordinates {(2.30278,1)(2.4495,1)(3,1)};
    \legend{}

\addplot[color=black,mark=triangle*]
    coordinates {(1.79868,412)(1.84891,9)(1.89933,1)(1.9498,1)(2,1)(2.10381,1)(2.15444,1)(2.23608,1)(2.30278,1)(2.4495,1)(3,1)};
    \legend{}

\addplot[color=blue,mark=square*]
    coordinates {(2.30278,1)(2.4495,1)(3,1)};
    \legend{}

\end{axis}

\end{tikzpicture}
}
\end{minipage}
\begin{minipage}[t]{0.49\textwidth}
\resizebox {0.99\textwidth} {!} {
\begin{tikzpicture}
\begin{axis}[
    axis y line*=left,
    title={Algorithm variants for \textsc{$4$-PVC}},
    xlabel={Branching factor $\beta$},
    ylabel={Running time in seconds},
    ymode=log,
    legend pos=north west,
    ymajorgrids=true,
    grid style=dashed,
]

\addplot[color=blue,mark=square*]
    coordinates {(3,1)(3.10381,1)(3.16229,1)(3.30279,1)(4,1)};
    \legend{}

\addplot[color=black,mark=triangle*]
    coordinates {(2.29299,799)(2.34314,143)(2.39348,42)(2.44373,11)(2.49425,5)(2.54831,2)(2.59868,1)(2.65898,1)(2.71443,1)(2.77846,1)(2.83848,1)(2.89512,1)(2.946,1)(3,1)(3.10381,1)(3.16229,1)(3.30279,1)(4,1)};
    \legend{}

\addplot[color=purple,mark=diamond*]
    coordinates {(3,1)(3.10381,1)(3.16229,1)(3.30279,1)(4,1)};
    \legend{}

\addplot[color=orange,mark=otimes*]
    coordinates {(2.36944,2375)(2.41948,181)(2.46967,41)(2.51985,13)(2.57013,5)(2.62087,3)(2.67389,1)(2.72675,1)(2.77846,1)(2.83848,1)(2.89512,1)(2.946,1)(3,1)(3.10381,1)(3.16229,1)(3.30279,1)(4,1)};
    \legend{}

\addplot[color=red,mark=x]
    coordinates {(3.03576,1)(3.10381,1)(3.16229,1)(3.30279,1)(4,1)};
    \legend{}

\end{axis}

\end{tikzpicture}
}
\end{minipage}
\begin{minipage}[t]{0.49\textwidth}
\resizebox {0.99\textwidth} {!} {
\begin{tikzpicture}
\begin{axis}[
    axis y line*=left,
    title={Algorithm variants for \textsc{$5$-PVC}},
    xlabel={Branching factor $\beta$},
    ylabel={Running time in seconds},
    ymode=log,
    legend pos=north west,
    ymajorgrids=true,
    grid style=dashed,
]

\addplot[color=black,mark=triangle*]
    coordinates {(2.84971,2552)(2.89977,929)(2.94986,409)(3,181)(3.05425,82)(3.10435,42)(3.15517,29)(3.20702,12)(3.25791,8)(3.30851,5)(3.36524,4)(3.41631,3)(3.47091,2)(3.52512,1)(3.58688,1)(3.64576,1)(3.70157,1)(3.7913,1)(3.873,1)(4,1)(4.06066,1)(4.23608,1)(5,1)};
    \legend{}

\addplot[color=orange,mark=otimes*]
    coordinates {(2.95731,3693)(3.00736,1132)(3.05739,620)(3.1074,216)(3.15739,169)(3.20953,88)(3.26004,36)(3.31041,13)(3.36524,6)(3.41631,4)(3.47091,2)(3.52405,2)(3.58688,1)(3.64576,2)(3.70157,1)(3.7913,1)(3.873,1)(4,1)(4.06066,1)(4.23608,1)(5,1)};
    \legend{}

\addplot[color=blue,mark=square*]
    coordinates {(3.64576,2)(3.70157,1)(3.7913,1)(3.873,1)(4,1)(4.06066,1)(4.23608,1)(5,1)};
    \legend{}

\addplot[color=red,mark=x]
    coordinates {(5,1)};
    \legend{}

\addplot[color=purple,mark=diamond*]
    coordinates {(4,1)(4.06066,1)(4.23608,1)(5,1)};
    \legend{}

\end{axis}

\end{tikzpicture}
}
\end{minipage}
\end{center}

\caption{Comparison of the progression of running times of different variants of our algorithm.}
\label{modifier-comparison-plots}
\end{figure}

In \autoref{modifiers-comparsion-table} and \autoref{modifier-comparison-plots} we compare the behavior of the different algorithm variants that have some parts of the algorithm turned off. The experiment was to obtain the best branching factor given the time limit of 90 minutes.

As you can see, with everything turned off (see NONE), the algorithm is severely crippled, i.e., in the case of 5-\textsc{PVC}, it cannot even do better than the trivial branching. Once we introduce the handmade rules, few nontrivial results emerge (see HR only). Also, the handmade rules play a crucial role for the cases of 4-\textsc{PVC} and 5-\textsc{PVC}, as without them, no progress is made (see FA, no HR).
Finally, the absence of \textit{Adjust} function does not seem to severely hinder the algorithm (see HR, DF only), but compared to the full algorithm (see FA), one can see that it significantly speeds up the computation which allows us to get better algorithms given our limited resources (see \autoref{modifier-comparison-plots}).

\section{Annotated Descriptions of the Obtained Algorithms}
\label{sec:proofs}

Next to the repository \url{https://github.com/generating-algorithms/generating-dpvc} containing the source code, we also provide a separate repository \url{https://github.com/generating-algorithms/generating-dpvc-data} with annotated descriptions of the obtained algorithms. These are basically logs of the successful computation paths taken by the algorithm.

The purpose of these is twofold.

First, as with any computer program, it is hard to get fully convinced about correctness of the generating procedure.
Then, this annotated description of the output algorithm is something that can be verified independently of the generating procedure.
In fact, we provide small python scripts to do so.

Furthermore, for algorithms with fewer branching rules this could be even made by hand---we provide a script to translate the machine-readable \texttt{.json} description into a human readable set of interlinked HTML pages.
However, we provide the HTML pages compressed and only for few very small algorithms, as even for them the HTML description (after decompression) takes several gigabytes.
The examples can be found in the above-mentioned repository, but to obtain an example quickly, we provide an explicit link \url{https://github.com/generating-algorithms/generating-dpvc-data/raw/master/5_3.0742_20/5_3.0742_20.proof_visualization.tar.xz}.

Second, the description also allows to implement the output algorithm in a much more efficient way.
Before we describe how, let us first delve in the structure of the description.

The description basically includes for each graph that appeared in the set $\gLbad$ during the course of the algorithm a page explaining the reason why it was removed from the set.
Hence, it starts with the graphs from \autoref{initial-list-observation}, called \emph{initial graphs}.
There are only two reason why a graph $H$ can be removed from $\gLbad$:
\begin{enumerate}[a)]
 \item Either it gave rise to a good branching rule (or was identified as handled by~$\A$) and was removed on line \ref{removal-to-good} of \autoref{fab-algorithm-pseudocode},
 \item or it was expanded on line \ref{expand-line-fab-algorithm-pseudocode} of the algorithm.
\end{enumerate}

In case b), the annotated description contains all 1-expansions of $H$, each on one line, sorted according to the neighborhood of the new vertex.
They are equipped with an information, whether the expansion was treated further (included in $\gLbad$) or whether it was an expansion of a graph $F$ already in $\gLgood$.
In the former case it contains a link to the page of the expansion and the appropriate isomorphism (as several expansions can give rise to only one graph).
In the latter case it contains the subgraph isomorphism proving that it is an expansion of $F$ and a link to the page of $F$.

For the case a), the page first also contains the 1-expansions.
For those that are expansion of graphs already in $\gLgood$ it contains the same information as above.
This allows to verify that the vertices in $R$ are obtained according to function \textit{Color}.
Having the set $R$ at hand, we either receive the information that the pair $(H,R)$ is handled by the handmade rules, or we proceed further to the branches of the rule.

In the section devoted to branches of the rule, for each subset $S \subseteq V(H)$ there is an information that
\begin{enumerate}
 \item $S$ is not a solution for $H$. This is certified by providing the vertices of a $P_d$ in $H \setminus S$.
 \item $S$ is a solution, but not a minimal one. This is certified by a set $S' \subsetneq S$ which is a solution.
 \item $S$ is a solution, but dominated by another one. This is certified by the other branch $B_d$ together with the sets $B^{\mathit{del}}$ and $R^*$ as of \autoref{def:dominated_branch}.
 \item $S$ was replaced by a different branch during adjustment, providing this new branch.
 \item $S$ is one of the actual branches of the branching rule.
\end{enumerate}
With this information it is easy to verify that the rule is correct and has the claimed branching factor.

The graphs in the description are ordered according to the order in which they were removed from $\gLbad$, hence it is easy to verify which graphs were already in $\gLgood$ when this graph was considered. With this information it is easy to verify the exhaustiveness of the set of the rules.

Now to implement the described algorithm, we first use some algorithm to find an occurrence $P$ of a $P_d$ in the input graph $G$. Now the graph $H = G[V(P)]$ must be among the initial graphs. If it was expanded, then we simply take any neighbor $w$ of $V(P)$ in $G$. Based on the neighbors of $w$ in $V(P)$ we follow the appropriate 1-expansion of $H$, letting $H = G[V(P) \cup \{w\}]$.
If $H$ has actually no neighbors in $G$, then it forms a small component and we find a solution for it by brute-force.
We repeat this as long as the current graph $H$ was expanded.
If the expansion was eliminated by some graph in $\gLgood$, then we continue with the corresponding subgraph from $\gLgood$.

If the current subgraph $H$ of $G$ yields a branching rule, then we check whether some vertex of $R$ has a neighbor outside $H$.
If this is not the case, then we can simply apply the rule.
If some vertex $v \in R$ has a neighbor $w$ outside $H$, then we follow the 1-expansion of $H$ by $w$, restricting ourselves to the corresponding subgraph from $\gLgood$.
As this way we always arrive at a rule with lower number, at some point we must arrive at a rule which will be applied.

The parameter \texttt{rule\_walk\_length} provided in the description of the algorithms captures the maximum number of graphs we have to visit before some rule is applied. In our results, this number amounts to dozens even for algorithms with hundreds of thousands or even millions of rules. As each of the described steps can be done in linear time, this provides an efficient way to apply the algorithm, independent of the actual number of rules.

\section{Future Research Directions}
\label{sec:conclusions}

We provided a framework to generate parameterized branching algorithms tailored for specific vertex deletion problems.
In comparison, the framework of Gramm et al.~\cite{GrammGHN04} is also suited for problems where the task is to either delete or even add \emph{edges} to the graph. We wonder whether some of our ideas can be translated to the edge setting.

While there are rather few studies on computer generated algorithms with provable worst-case running time upper bounds, there are quite some papers that use computer aided \emph{analysis} of algorithms.
In particular, the \emph{Measure \& Conquer} approach, introduced by Fomin et al.~\cite{FominGK09}, is popular especially for moderately exponential algorithms~\cite{LokshtanovSS14,RooijB11,RooijB12,XiaoN17}.
Here the idea is to use simple rules, while measuring the progress not only based on the number of vertices resolved, but also on how favorably the remaining graph is structured, e.g., how many vertices of rather low degree are present.
The hope is to capture that some unfavorable branching significantly improves the structure so that a favorable branching appears subsequently. To accomplish this, the analysis of a single rule is often split into many cases, based, e.g., on the degrees of the vertices involved. The computer is then used to optimize the values assigned to favorable structures so as to prove the lowest possible worst-case running time upper bound.
Other approaches trying to amortize between the rules with bad branching factors and those with good branching factors include \emph{branching potential}~\cite{Gaspers10} or \emph{labeled search trees}~\cite{ChenKX05}. See also Fernau and Raible~\cite{FernauR09} for an older survey of the topic.

It may seem interesting to combine the automated generation framework with a computer assisted analysis of the algorithm. However, first, it seems that the computer assisted analysis still requires a non-trivial amount of human intervention, e.g., in design of the measure and cases to be distinguished by the computer. Therefore it seems to be limited to algorithms with few branching rules and does not scale to thousands of rules. Second, the favorable structure we gain, if it can be captured in an automated manner at all, is then exploited in the immediate neighborhood of the finished branching to gain the advantage. Hence, we might possibly as well create a single branching rule encompassing both the structures and ``amortize within the rule''. Of course, many variants of such a rule would be necessary. This is the approach already prevalent in our framework. However, the sizes of the rules necessary might be beyond the reach of our implementation. The question is whether some transfer of ``branching potential'' or some other kind of advantage can be explicitly included in the construction of the rules in order to enable this advanced analysis.

Finally, an obvious open question is whether there are, e.g., some handmade rules that would help our algorithm generate a faster algorithm for \textsc{Vertex Cover} (2-PVC). The fastest known algorithms of Chen, Kanj, and Xia~\cite{ChenKX10} and Harris and Narayanaswamy~\cite{HarrNaraFastVC} are rather complex to both analyze (both from the running time and correctness perspective) and implement. We made some experiments with the \emph{struction} and \emph{vertex-domination} rules from~\cite{ChenKX10}, but these did not seem to improve the performance of the generating algorithm.





\bibliographystyle{splncs04}
\bibliography{main}

\clearpage
\appendix

\appendixText

\end{document}